\documentclass[english,11pt,a4paper]{article}
\usepackage[T1]{fontenc}
\usepackage[latin9]{inputenc}
\usepackage{dsfont}
\usepackage{amsmath}
\usepackage{amsthm}
\usepackage{amssymb}
\usepackage{graphicx}

\makeatletter
\numberwithin{equation}{section}

\usepackage{jheppub}
\usepackage{orcidlink}
\usepackage{fontawesome5}
\allowdisplaybreaks

\newcommand{\chipck}{ {\chi}'^{*}_k}
\usepackage{tensor}

\DeclareDocumentCommand{\GitHub}{s}{%
	\IfBooleanTF{#1}{%
		\faGithub~\url{https://github.com/xunjiexu/Unified-Bogoliubov.git}%
	}{%
		\href{https://github.com/xunjiexu/Unified-Bogoliubov.git}{\faGithub}%
	}%
}

\makeatother

\usepackage{babel}
\begin{document}
\title{A Unified Bogoliubov Approach to Primordial Gravitational
Waves: From Inflation to Reheating}

\abstract{We present an effective numerical method that can be used
to straightforwardly calculate the full spectrum of primordial gravitational
waves produced during inflation and reheating. Our method is based
on the Bogoliubov approach with several key improvements to overcome
its shortcomings  such as numerical errors at high frequencies and
issues with tachyonic modes. We also present a few useful analytical
examples from which one can gain crucial insights into the numerical
errors. The improved method allows us to demonstrate that anharmonicity
of inflaton oscillations can leave interesting fingerprints on the
high-frequency part of the GW spectrum.  Our numerical code is publicly
available on GitHub~\GitHub.

}

\author[a]{Yubing Wang \orcidlink{0009-0008-2577-5247}} 
\author[b,c]{Quan-feng Wu \orcidlink{0000-0002-5716-5266}}
\author[b]{and Xun-Jie Xu \orcidlink{0000-0003-3181-1386}}
\affiliation[a]{Department of Physics and Astronomy, University of Bonn, Bonn 53115, Germany}
\affiliation[b]{Institute of High Energy Physics, Chinese Academy of Sciences, Beijing 100049, China}
\affiliation[c]{Kaiping Neutrino Research Center, Kaiping 529386, China}
\emailAdd{s56ywang@uni-bonn.de} 
\emailAdd{wuquanfeng@ihep.ac.cn} 
\emailAdd{xuxj@ihep.ac.cn} 
\preprint{\today}
\maketitle

\section{Introduction}

\begin{figure}

\centering

\includegraphics[width=0.85\textwidth]{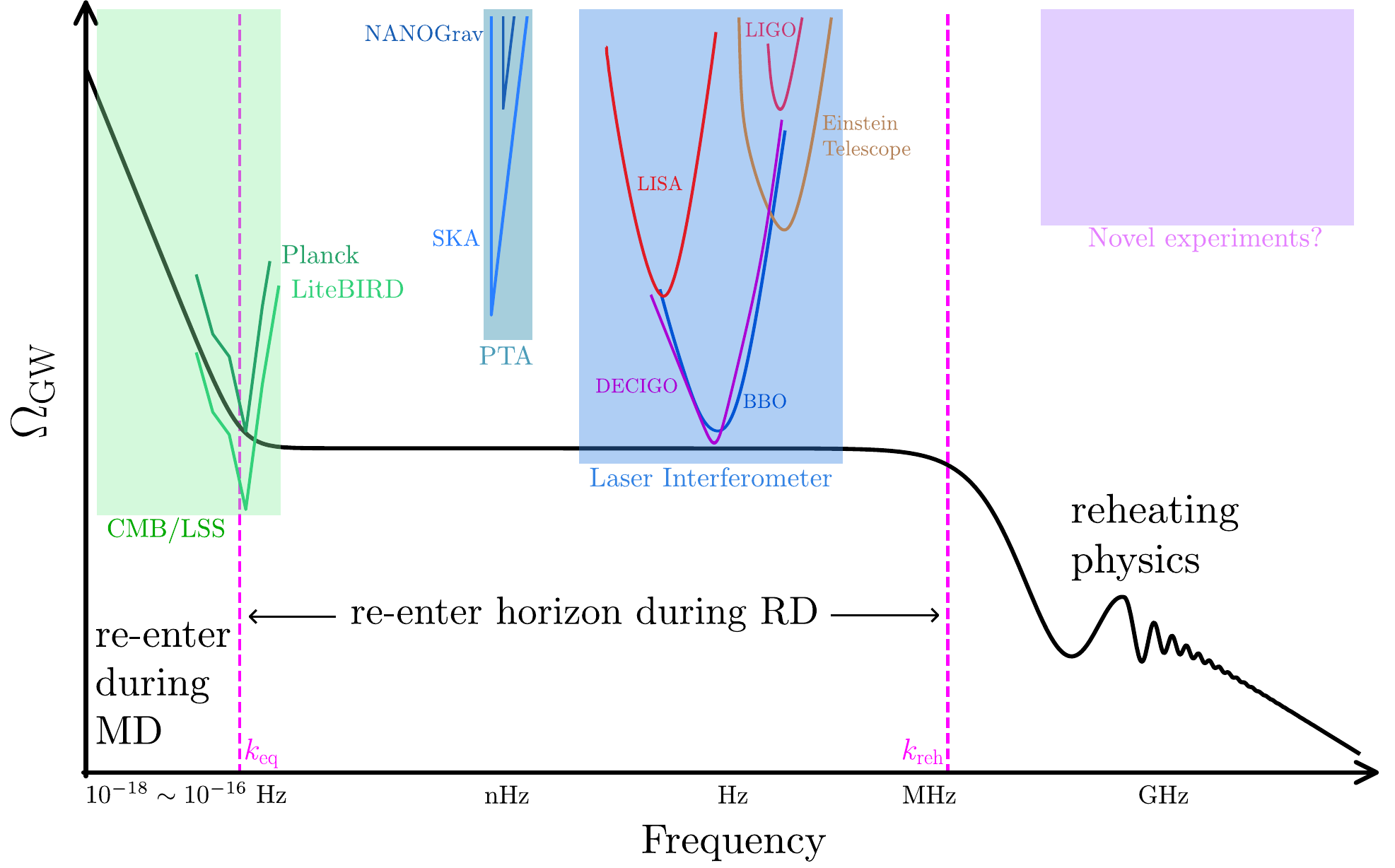}\caption{\label{fig:broad-spectrum} Schematic illustration (not to scale)
of the broad GW spectrum predicted by cosmic inflation (solid black
line) and the prospects for future experimental probes. }
\end{figure}

Cosmic inflation generically predicts the presence of  primordial
gravitational waves (GWs) spanning an extraordinarily broad range
of frequencies---see Ref.~\cite{Caprini:2018mtu} for a comprehensive
review. The broad GW spectrum, together with many relevant experimental
probes, is schematically illustrated in Fig.~\ref{fig:broad-spectrum},
where the spectrum extends from ultra-low frequencies accessible
indirectly via cosmic microwave background (CMB) observations including
Planck~\cite{Planck:2018jri} and future missions like LiteBIRD~\cite{LiteBIRD:2022cnt},
through the nano-Hz regime targeted by pulsar timing arrays (PTAs)
including NANOGrav~\cite{NANOGrav:2023gor} and SKA~\cite{Janssen:2014dka},
up to the Hz band relevant for space-based interferometers such as
LISA~\cite{LISA:2017pwj}, BBO~\cite{Crowder:2005nr,Corbin:2005ny,Harry:2006fi},
and DECIGO~\cite{Kawamura:2011zz}.   At even higher frequencies,
above MHz or GHz, there has been growing interest in novel detection
ideas~\cite{Domcke:2020yzq,Berlin:2021txa,Domcke:2022rgu,Herman:2022fau,Vagnozzi:2022qmc,Bringmann:2023gba,Kahn:2023mrj,He:2023xoh,Liu:2023mll,Kanno:2023whr,Valero:2024ncz,Pappas:2025zld}
aimed at probing this regime---see Refs.~\cite{Aggarwal:2020olq,Aggarwal:2025noe}
for reviews. 

The high-frequency part of this spectrum is particularly intriguing.
Unlike the nearly scale-invariant plateau generated during inflation,
the spectrum at high frequencies can deviate significantly from scale
invariance and exhibit nontrivial features reflecting the dynamics
of reheating.  Hence, this regime offers a unique observational window
into reheating physics, potentially revealing information inaccessible
through lower-frequency probes.

Recently, Ref.~\cite{Choi:2024ilx} studied in this regime the production
of GWs through a minimal mechanism: inflaton annihilation. Assuming
only gravitational interactions of the inflaton ($\phi$), one expects
a considerable number of gravitons ($h$) to be produced after inflation
from the dense environment of inflatons via $\phi\phi\to hh$. Ref.~\cite{Choi:2024ilx}
demonstrates that, depending on the equation of state during reheating,
the resulting GW spectrum may exhibit different shapes; if the inflaton
potential around its minimum approximates to $V\propto\phi^{\alpha}$,
the spectrum scales as $\nu^{\frac{4\alpha-7}{\alpha-4}}$ with $\nu$
the GW frequency.    Beyond this minimal scenario, a growing body
of literature~\cite{Nakayama:2018ptw,Huang:2019lgd,Barman:2023ymn,Barman:2023rpg,Bernal:2023wus,Kanemura:2023pnv,Tokareva:2023mrt,Xu:2024fjl,Bernal:2024jim,Barman:2024htg,Hu:2024awd,Inui:2024wgj,Jiang:2024akb,Bernal:2025lxp,Xu:2025wjq,Das:2025cqs,Cheng:2025gyh,Bernal:2026dsu}
 has investigated additional GW production channels, including bremsstrahlung
associated with inflaton decay and scattering processes among inflaton
decay products. 

In this work, we concentrate on the minimal production of GWs in the
inflationary framework. Our primary goal is to establish a unified
approach that can be used to straightforwardly and effectively obtain
the full spectrum without imposing any assumptions (e.g., the slow-roll
approximation or matter domination during reheating). Here the full
spectrum covers the nearly scale-invariant part arising from inflation,
certain power-law forms from reheating, and the transition between
them. As we will show, this is numerically challenging, but assisted
with crucial analytical insights, it can be achieved very successfully.
In particular, using this method, we demonstrate that anharmonicity
of inflaton oscillations can leave interesting fingerprints on the
high-frequency GW spectrum (illustrated by the wiggles on the spectrum
in Fig.~\ref{fig:broad-spectrum}). 

 While many calculations of GW production during reheating in the
above references are formulated within the Boltzmann approach,  we
adopt the Bogoliubov approach, which interprets particle production
as an effect of dynamical spacetime and has also been used for computing
primordial GWs~\cite{Ema:2015dka,Ema:2016hlw,Ema:2020ggo,Pi:2024kpw,Mudrunka:2026kgm,Wang:2026ule}.
 Both approaches have their respective shortcomings: the Boltzmann
approach can only be applied to scenarios where particles can be well
defined with their de Broglie or Compton wavelengths well limited
within the horizon, such that one can use Minkowski-spacetime quantum
field theory (QFT) locally. Hence the Boltzmann approach is, by construction,
more suitable for high-frequency GW calculations. And it becomes invalid
when a relevant length scale becomes comparable to or exceeds the
horizon.  The Bogoliubov approach, on the other hand, often requires
solving differential equations numerically and is known to often cause
numerical errors when computing high-frequency modes, though in principle
it is widely applicable to any frequency.  Several recent studies
have shown that the two approaches in certain cases lead to consistent
results~\cite{Kaneta:2022gug,Chakraborty:2025zgx,Wang:2026ule}.

This work extends and improves the conventional Bogoliubov approach
in several respects. First, we identify potentially large cancellations
that may cause numerical errors in the Bogoliubov approach and propose
a practically useful parametrization to effectively circumvent large
cancellations. The Wronskian condition plays a crucial role in circumventing
large cancellations. Second, we employ the adiabaticity parameter
to determine the effective production epoch. This allows us to avoid
unnecessary evolution of differential equations in irrelevant time
periods, substantially increasing the numerical efficiency.  Finally,
we incorporate a technique which we refer to as \emph{ultra-violet
(UV) smoothing} to avoid UV noise, which would otherwise interfere
with high-frequency physical effects (such as the aforementioned wiggles
on the spectrum). Our improved method may be useful for studies encountering
similar numerical issues  in the Bogoliubov approach. Our numerical
code is publicly available on GitHub \GitHub. 

Throughout this work, we adopt the $(+,-,-,-)$ sign convention. The
scale factor and Hubble parameter are denoted by $a$ and $H=\dot{a}/a$,
respectively. We use $\eta$ and $t$ to denote the conformal and
cosmic (physical) time. The derivatives with respect to $t$ and $\eta$
are denoted by dots (e.g., $\dot{a}\equiv da/dt$) and primes ($a'\equiv da/d\eta$).
We also use $M_{P}\approx2.43\times10^{18}$ GeV for the reduced Planck
mass, and assume that at sufficiently high temperatures the effective
number of relativistic degrees of freedom $g_{\star}$ is $106.75$.

The structure of this paper is organized as follows. In Sec.~\ref{sec:formulation},
we briefly review the Bogoliubov formalism to be used for the GW spectrum
calculation. In Sec.~\ref{sec:Analytical-solutions}, we present
several analytical solutions and highlight the analytical insight
into the issues of large cancellations and UV divergences. This facilitates
the  development of our improved numerical method for the Bogoliubov
approach, which is elaborated in Sec.~\ref{sec:D-parametrization}.
In Sec.~\ref{sec:Result}, we apply our numerical method to two specific
inflation models which are phenomenologically viable, and demonstrate
that they lead to distinct high-frequency features on the GW spectrum.
 Finally, we conclude in Sec.~\ref{sec:conclusion} and relegate
some details to the appendices.

\section{The Bogoliubov approach\label{sec:formulation}}

The Bogoliubov approach to particle production in dynamical spacetime
has been widely applied to the so-called Cosmological Gravitational
Particle Production (CGPP)\,---\,see Ref.~\cite{Kolb:2023ydq}
for a review. In an expanding universe where the curvature may vary
rapidly, the mode functions in QFT associated with the Bunch-Davies
vacuum (i.e., the vacuum in the infinite past) may not match the vacuum
state in future asymptotically flat spacetime, leading to particle
production. In spirit, this resembles how a ground-state wavefunction
in quantum mechanics evolves when a time-dependent potential is introduced
into the Schr\"odinger equation. 

A quantitative study of this effect for a massless\footnote{For a massive species with mass $m$, one adds $a^{2}m^{2}$ to $k^{2}$
in Eq.~\eqref{eq:}.} particle species (e.g., the graviton) requires solving the following
equation of the mode function~\cite{Kolb:2023ydq}:
\begin{equation}
\chi_{k}''(\eta)+\left[k^{2}-\mu^{2}\left(\eta\right)\right]\chi_{k}(\eta)=0\thinspace,\label{eq:}
\end{equation}
with 
\begin{equation}
\mu^{2}\left(\eta\right)\equiv\frac{a''}{a}=-\frac{1}{6}a^{2}R\thinspace.\label{eq:-1}
\end{equation}
Here $\chi_{k}(\eta)$ is the mode function, which is a function of
the comoving momentum $k$ and the conformal time $\eta$. We refer
to Ref.~\cite{Kolb:2023ydq} for detailed discussions on the physical
interpretation of the mode function and how it is related to the particle
number.   The Ricci scalar can be calculated as follows:
\begin{equation}
R=-6\frac{a''}{a^{3}}=-6\left(\dot{H}+2H^{2}\right)=\sum_{i}\frac{3p_{i}-\rho_{i}}{M_{P}^{2}}\thinspace,\label{eq:-50}
\end{equation}
where $p_{i}$ and $\rho_{i}$ denote the pressure and energy density
of species $i$. Eq.~\eqref{eq:-50} implies $R\approx0$ in radiation
domination (RD) and $R\approx-12H^{2}$ in the slow-roll (SR) epoch.
Correspondingly, we have $\mu^{2}\approx2/\eta^{2}$ and $\mu^{2}\approx0$
during SR and RD epochs, respectively. In matter domination (MD),
$\mu^{2}\approx2/(\eta-C)^{2}$ with a constant $C$ to be fixed by
the beginning of the MD epoch. In Appendix~\ref{sec:relations} we
briefly summarize these relations.

In the infinite past, the Bunch-Davies vacuum requires the following
asymptotical behavior:
\begin{equation}
\chi_{k}\to\frac{1}{\sqrt{2k}}e^{-ik\eta}\ \ \text{for}\ \ \eta\to-\infty\thinspace.\label{eq:-23}
\end{equation}
If a solution to Eq.~\eqref{eq:} is obtained and satisfies Eq.~\eqref{eq:-23},
then the particle production seen in future asymptotically flat spacetime
is given by~\cite{Kolb:2023ydq}
\begin{equation}
f\left(k\right)=\lim_{\eta\to+\infty}\frac{\omega_{k}}{2}\left|\chi_{k}\right|^{2}+\frac{1}{2\omega_{k}}\left|\chi_{k}'\right|^{2}-\frac{1}{2}\thinspace,\label{eq:-3}
\end{equation}
where $\omega_{k}\equiv\sqrt{k^{2}-\mu^{2}}$ and $f(k)$ is the
phase space distribution function widely used in the Boltzmann approach.
 Note that here we are using the comoving momentum $k$, not the
physical momentum, which is related to $k$ by
\begin{equation}
k_{{\rm phy}}=k/a\thinspace.\label{eq:-37}
\end{equation}

In practice, it is useful to parametrize the mode function as \cite{Kofman:1997yn}\footnote{Our $\alpha_{k}$ and $\beta_{k}$ correspond to $\tilde{\alpha}_{k}$
and $\tilde{\beta}_{k}$  in Ref.~\cite{Kolb:2023ydq}. }
\begin{equation}
\chi_{k}=\alpha_{k}\frac{e^{-i\Phi_{k}}}{\sqrt{2\omega_{k}}}+\beta_{k}\frac{e^{i\Phi_{k}}}{\sqrt{2\omega_{k}}}\thinspace,\label{eq:-7}
\end{equation}
where $\Phi_{k}(\eta)=\int^{\eta}\omega_{k}(\tilde{\eta})d\tilde{\eta}$.
Then Eq.~\eqref{eq:} is equivalent to 
\begin{align}
\alpha'_{k} & =\frac{\omega'_{k}}{2\omega_{k}}\beta_{k}e^{2i\Phi_{k}}\thinspace,\label{eq:-8}\\
\beta'_{k} & =\frac{\omega'_{k}}{2\omega_{k}}\alpha_{k}e^{-2i\Phi_{k}}\thinspace.\label{eq:-9}
\end{align}
When $k$ is very large ($k^{2}\gg\mu^{2}$) and $\mu^{2}$ varies
slowly compared to the fast oscillation of $\chi_{k}$, one expects
that $\chi_{k}$ should evolve adiabatically (similar to adiabatic
processes in quantum mechanics). In the adiabatic limit, $\omega'_{k}/\omega_{k}$
in Eqs.~\eqref{eq:-8} and \eqref{eq:-9} is much smaller than $\omega_{k}$,
which is responsible for the oscillation of $\chi_{k}$. Hence $\alpha_{k}$
and $\beta_{k}$ in this limit evolve much more mildly than $\chi_{k}$,
significantly facilitating numerical calculations. 

Using Eq.~\eqref{eq:-3} and the Wronskian condition
\begin{equation}
\chi_{k}\chipck-\chi_{k}^{*}\chi'_{k}=i\thinspace,\label{eq:-10}
\end{equation}
it is straightforward to verify that 
\begin{equation}
f=\lim_{\eta\to+\infty}|\beta_{k}|^{2}\thinspace.\label{eq:-11}
\end{equation}
Given the simple relation in Eq.~\eqref{eq:-11} and the aforementioned
advantages in numerical calculations,  solving the differential equations
of $(\alpha_{k},\ \beta_{k})$ instead of $\chi_{k}$ is a widely
adopted approach in the literature. However, this approach requires
that $\omega_{k}^{2}>0$, otherwise a singularity would arise from
the denominators in Eqs.~\eqref{eq:-8} and \eqref{eq:-9} when $\omega_{k}^{2}$
crosses zero. Therefore, one usually avoids using the above $\alpha$-$\beta$
parametrization in cases where tachyonic modes ($\omega_{k}^{2}<0$)
could occur~\cite{Mudrunka:2026kgm}. 

Once the production completes, the subsequent evolution of $f$ is
simple if the produced particles free-stream in the universe, which
is the case for gravitons. In this case, $f$ as a function of the
comoving momentum $k$ remains unchanged in the subsequent evolution.\footnote{See, however, Refs.~\cite{Boyle:2005se,Saikawa:2018rcs} for discussions
on minor effects that can modify this. } If expressed in terms of the physical momentum $k_{{\rm phy}}$,
the only change is the red-shift of $k_{{\rm phy}}$ due to the Hubble
expansion. 

For gravitons produced in the early universe and red-shifted to the
present, one can evolve $f$ towards its final (stable) value and
use the following formula to compute their contribution to the cosmic
energy budget~\cite{Xu:2025wjq}:
\begin{equation}
\Omega_{{\rm GW}}\equiv\frac{1}{\rho_{c}}\frac{d\rho_{{\rm GW}}}{d\ln k_{{\rm phy}}}\approx3.7\times10^{-13}h^{-2}\left(\frac{\nu}{\text{GHz}}\right)^{4}f\thinspace.\label{eq:-17}
\end{equation}
Here $\rho_{c}\simeq1.05\times10^{-5}h^{2}\,\text{GeV}/\text{cm}^{3}$
is the critical energy density, $\rho_{\text{GW}}$ is the energy
density of GWs, and $\nu$ is the GW frequency, related to $k=k_{{\rm phy}}a$
by
\begin{equation}
\frac{\nu}{\text{GHz}}\approx45.9\left(\frac{g_{\star}}{106.75}\right)^{-\frac{1}{12}}\frac{k}{\rho_{r}^{1/4}a}\thinspace,\label{eq:-18}
\end{equation}
where $\rho_{r}$ is the energy density of radiation, evaluated in
the radiation-dominated epoch, during which $\rho_{r}^{1/4}a$ in
the denominator is a constant.  

\section{Analytical solutions \label{sec:Analytical-solutions}}

There are a few forms of $\mu^{2}(\eta)$ that allow one to solve
Eq.~\eqref{eq:} analytically. In this section, we present several
analytical solutions and discuss their physical implications. These
analytical solutions are important for three main reasons. 
\begin{itemize}
\item First, with analytical solutions that are approximately valid in certain
epochs (e.g., SR, MD, and RD),  one can construct piecewise solutions
to approximate the exact form of $\chi_{k}(\eta)$.  In particular,
for $\chi_{k}$ evolving from SR to RD with instantaneous reheating,
one can use analytical solutions to readily obtain the nearly scale-invariant
GW spectrum predicted by SR inflation.  
\item Second, as we will see below, the calculation of $f(k)$ from $\chi_{k}(\eta)$
often involves large cancellations, which is partially the cause of
large numerical errors  in the Bogoliubov approach. The analytical
solutions allow us to explicitly examine how such cancellations emerge.
\item Third, some analytical solutions demonstrate that UV divergences arise
if $\mu^{2}$ is discontinuous. Although discontinuities in $\mu^{2}$
are unphysical, they may appear in numerical calculations with finite
precision. Besides, continuous but non-smooth $\mu^{2}$ may result
in UV noise, which typically behave as unphysical oscillations. From
the analytical solutions, one can gain important insight into the
UV divergences and UV noise. This can be helpful for numerical calculations.
\end{itemize}
   When $\mu^{2}$ takes one of the following forms, there are
known analytical solutions:
\begin{align}
\mu^{2}=\frac{2}{\eta^{2}} & \ \ \Rightarrow\ \ \chi_{k}=\frac{1}{\sqrt{2k}}\left(1\mp\frac{i}{k\eta}\right)e^{\mp ik\eta}\thinspace;\label{eq:-19}\\
\mu^{2}=0 & \ \ \Rightarrow\ \ \chi_{k}=\frac{1}{\sqrt{2k}}e^{\mp ik\eta}\thinspace;\label{eq:-20}\\
\mu^{2}=\frac{\lambda}{\eta^{2}} & \ \ \Rightarrow\ \ \chi_{k}=\sqrt{\eta}J_{\frac{1}{2}\sqrt{4\lambda+1}}(k\eta)\ \text{or}\ \sqrt{\eta}Y_{\frac{1}{2}\sqrt{4\lambda+1}}(k\eta)\thinspace;\label{eq:-21}\\
\mu^{2}=\lambda\cos(2m\eta) & \ \ \Rightarrow\ \ \chi_{k}=\mathds{C}\left(\frac{k^{2}}{m^{2}},\frac{\lambda}{2m^{2}},\eta m\right)\ \text{or}\ \mathds{S}\left(\frac{k^{2}}{m^{2}},\frac{\lambda}{2m^{2}},\eta m\right).\label{eq:-22}
\end{align}
Here $J$ and $Y$ are the Bessel functions of the first and second
kinds; and $\mathds{C}$ and $\mathds{S}$ are the even and odd Mathieu
functions, respectively. In the SR or MD epoch, $\mu^{2}$ can be
approximated by $\frac{2}{\eta^{2}}$ (up to a constant shift in $\eta$,
see Appendix~\ref{sec:relations}) with $\eta<0$ or $\eta>0$, respectively.
 The case of $\mu^{2}=0$ corresponds to the epoch of RD.  At the
end of inflation, a short period of $w\equiv p_{\phi}/\rho_{\phi}\approx1$
may occur, known as the kination domination (KD), corresponding to
$\mu^{2}=\lambda/\eta^{2}$ with $\lambda=-1/4$. The last case, $\mu^{2}=\lambda\cos(2m\eta)$,
can be used to qualitatively understand the production of particles
from an oscillatory spacetime background, in particular, the feature
of resonant production. 

\subsection{SR+RD (Instantaneous reheating) \label{subsec:SR+RD}}

Let us first consider the following $\mu^{2}$ function, which can
be physically related to the SR evolution followed by RD with instantaneous
reheating:
\begin{equation}
\mu^{2}\left(\eta\right)=\begin{cases}
\frac{2}{\eta^{2}} & \eta\leq\eta_{1}\\
0 & \eta>\eta_{1}
\end{cases}.\label{eq:-24}
\end{equation}
Note that here we require $\eta_{1}<0$, as the SR process should
end before $2/\eta^{2}$ reaches the singularity at $\eta=0$. 

By combining Eqs.~\eqref{eq:-19} and \eqref{eq:-20} and imposing
continuity conditions on $\chi_{k}$ and $\chi'_{k}$ at $\eta=\eta_{1}$,
it is straightforward to obtain the analytical solution:
\begin{equation}
\chi_{k}(\eta)=\begin{cases}
\frac{1}{\sqrt{2k}}\left(1-\frac{i}{k\eta}\right)e^{-ik\eta} & \eta\leq\eta_{1}\\
\left(1-\frac{i}{k\eta_{1}}-\frac{1}{2k^{2}\eta_{1}^{2}}\right)\frac{e^{-ik\eta}}{\sqrt{2k}}+\frac{e^{-2ik\eta_{1}}}{2k^{2}\eta_{1}^{2}}\cdot\frac{e^{+ik\eta}}{\sqrt{2k}} & \eta>\eta_{1}
\end{cases}\thinspace.\label{eq:-25}
\end{equation}
Substituting the solution into Eq.~\eqref{eq:-3}, we can compute
$f(k)$.  Here we would like to present explicitly the contribution
of each term in Eq.~\eqref{eq:-3}:
\begin{align}
\frac{k}{2}\left|\chi_{k}\right|^{2} & =\frac{1}{4}+\frac{1}{8k^{4}\eta_{1}^{4}}+A\cos\left[2k(\eta-\eta_{1})\right]-B\sin\left[2k(\eta-\eta_{1})\right],\label{eq:-35}\\
\frac{1}{2k}\left|\chi_{k}'\right|^{2} & =\frac{1}{4}+\frac{1}{8k^{4}\eta_{1}^{4}}-A\cos\left[2k(\eta-\eta_{1})\right]+B\sin\left[2k(\eta-\eta_{1})\right],\label{eq:-36}
\end{align}
where $A=\frac{2k^{2}\eta_{1}^{2}-1}{8k^{4}\eta_{1}^{4}}$ and $B=\frac{k\eta_{1}}{4k^{4}\eta_{1}^{4}}$.
Obviously one can see that the oscillatory parts of Eqs.~\eqref{eq:-35}
and ~\eqref{eq:-36} cancel out, and the two $1/4$ factors cancel
out with $-1/2$ in Eq.~\eqref{eq:-3}, leading to
\begin{equation}
f(k)=\frac{1}{4k^{4}\eta_{1}^{4}}\thinspace.\label{eq:-9-1}
\end{equation}
The above cancellations imply that Eq.~\eqref{eq:-3} is not suitable
for numerical calculation of $f(k)$ at high frequencies. For $k\eta_{1}\gg1$,
the term $\frac{1}{8k^{4}\eta_{1}^{4}}$ in Eq.~\eqref{eq:-35} or
\eqref{eq:-36} becomes highly suppressed in comparison with the overall
magnitude of $\frac{k}{2}\left|\chi_{k}\right|^{2}$ or $\frac{1}{2k}\left|\chi_{k}'\right|^{2}$.
Since numerical solutions  of the differential equation always contain
small numerical inaccuracies, the influence of such inaccuracies on
the resulting $f$ calculated via Eq.~\eqref{eq:-3} would be substantially
amplified.

Assuming a constant Hubble expansion rate during SR, we can use $\eta_{1}=-1/H_{I}a_{e}$,
where $a_{e}$ is the scale factor at the end of inflation and $H_{I}$
is the constant Hubble rate, to rewrite Eq.~\eqref{eq:-9-1} as
\begin{equation}
f=\frac{H_{I}^{4}a_{e}^{4}}{4k_{{\rm phy}}^{4}a^{4}}\thinspace.\label{eq:-10-1}
\end{equation}

According to Eq.~\eqref{eq:-17}, the above result implies the following
scale-invariant GW spectrum:
\begin{equation}
\Omega_{{\rm GW}}\simeq2.3\times10^{-16}\left(\frac{H_{I}}{10^{14}\ \text{GeV}}\right)^{2}\left(\frac{g_{\star}}{106.75}\right)^{-\frac{1}{3}}.\label{eq:-26}
\end{equation}

The above calculation raises a potential issue in the UV limit~\cite{Pi:2024kpw}:
the resulting scale-invariant spectrum formally extends to arbitrarily
large $k$, implying that the total energy density of GWs is UV divergent
($\rho_{{\rm GW}}\propto\int^{\infty}fkd^{3}\mathbf{k}\propto\int^{\infty}k^{-1}dk$
diverges logarithmically when $k\to\infty$)! 

One may question whether this behavior is physically meaningful. Physically,
it is well understood that modes with sufficiently large $k$ never
exit the horizon. So one may expect that such modes  are barely affected
by the background evolution with $\mu^{2}\ll k^{2}$, retaining their
original form associated with the Bunch-Davies vacuum. Indeed, the
second branch of Eq.~\eqref{eq:-25} in the limit of $k\eta_{1}\to\infty$
reduces to $e^{-ik\eta}/\sqrt{2k}$, up to terms suppressed by $(k\eta_{1})^{-1}$
or $(k\eta_{1})^{-2}$. 

The differential equation under consideration is analogous to a 1D
scattering problem in quantum mechanics. Consider how a wavefunction
in quantum mechanics is modulated by an effective potential. In the
regime where the potential is shallow ($\mu^{2}\ll k^{2}$), its influence
on the wavefunction is typically negligible. In the asymptotic limit
$k^{2}/\mu^{2}\to\infty$, one therefore expects the solution to approach
a free wave, effectively insensitive to the presence of the potential.
This expectation, however, is not universally valid. A notable exception
arises when the potential exhibits rapid spatial variation. In such
cases, even very energetic wavefunctions ($k^{2}\gg\mu^{2}$) may
experience a non-negligible influence from the potential. Consequently,
the naive argument that large-$k$ wavefunctions are unaffected by
the potential must be treated with care, particularly in situations
where the potential contains sharp features or strong gradients.

Back to the cosmological problem considered here, the UV divergence
arises from the sharp cut-off introduced in Eq.~\eqref{eq:-24}. In
a realistic cosmological model, the reheating process is always non-instantaneous
and cannot be approximated by the sharp cut-off. If the reheating
process takes finite time $\delta\eta$, then modes with $k\gg1/\delta\eta$
(see also the discussion on the adiabaticity parameter in Sec.~\ref{sec:D-parametrization})
should lead to suppressed contributions to $\Omega_{{\rm GW}}$. Appendix~\ref{sec:Suppression-of-UV}
demonstrates exactly this suppression. 

The UV divergence issue discussed above has an important implication
for numerical calculations: one should avoid unphysical non-smoothness
of the $\mu^{2}$ function obtained numerically, otherwise it may
lead to UV divergences or UV noise in the resulting GW spectrum.

\subsection{SR+MD+RD}

At the end of inflation, the inflaton field may oscillate at the bottom
of the potential. If the bottom is approximately quadratic, then it
corresponds to a MD phase. Eventually this MD phase will be replaced
by RD due to, e.g., inflaton decay. This is one of the most widely
studied scenarios for non-instantaneous reheating. Motivated by this
scenario, we consider the following $\mu^{2}$ function:
\begin{equation}
\mu^{2}\left(\eta\right)=\begin{cases}
\frac{2}{\eta^{2}} & \eta\leq\eta_{1}\\
\frac{2}{\left(\eta-2\eta_{1}\right)^{2}} & \eta_{1}<\eta\leq\eta_{2}\\
0 & \eta_{2}<\eta
\end{cases}\thinspace.\label{eq:-28}
\end{equation}
Compared to Eq.~\eqref{eq:-24}, it contains an extra part in $\eta_{1}<\eta\leq\eta_{2}$
corresponding to MD with decreasing $\mu^{2}$. Note that here $\eta_{1}$
is still negative, while $\eta_{2}$ can be positive or negative.
We have added a constant $-2\eta_{1}$ (positive) to $\eta$ in one
of the denominators such that $\mu^{2}$ is continuous at $\eta_{1}$.

In Eq.~\eqref{eq:-28}, we assume that the universe transitions from
MD to RD at $\eta=\eta_{2}$. Hence we view it as the end of reheating
and denote the corresponding scale factor at this moment by $a_{\text{rh}}$.
Then the cosmic expansion during the MD phase is given by 
\begin{equation}
\frac{a_{\text{rh}}}{a_{e}}=\frac{(\eta_{2}-2\eta_{1})^{2}}{\eta_{1}^{2}}\thinspace.\label{eq:-34}
\end{equation}

Similar to the calculation in Sec.~\ref{subsec:SR+RD}, we can solve
Eq.~\eqref{eq:} with $\mu^{2}$ given by Eq.~\eqref{eq:-28} and
obtain
\begin{equation}
\chi_{k}(\eta)=\begin{cases}
\frac{1}{\sqrt{2k}}\left(1-\frac{i}{k\eta}\right)e^{-ik\eta} & \eta\leq\eta_{1}\\
C_{1}\left[1-\frac{i}{k(\eta-2\eta_{1})}\right]\frac{e^{-ik\eta}}{\sqrt{2k}}+C_{2}\left[1+\frac{i}{k(\eta-2\eta_{1})}\right]\frac{e^{ik\eta}}{\sqrt{2k}} & \eta_{1}<\eta\leq\eta_{2}\\
\alpha\frac{e^{-ik\eta}}{\sqrt{2k}}+\beta\frac{e^{ik\eta}}{\sqrt{2k}} & \eta_{2}<\eta
\end{cases}\thinspace,\label{eq:-25-1}
\end{equation}
with 
\begin{align}
C_{1} & =\frac{i-2k\eta_{1}-2ik^{2}\eta_{1}^{2}+k^{3}\eta_{1}^{3}}{k^{3}\eta_{1}^{3}}\thinspace,\ \ \ C_{2}=\frac{ie^{-2ik\eta_{1}}}{k^{3}\eta_{1}^{3}}\thinspace,\label{eq:-29}\\
\alpha & =\frac{e^{2ik\eta_{2}}}{2\kappa^{2}}C_{2}+z_{\kappa}C_{1}\thinspace,\ \ \ \ \ \ \ \beta=\frac{e^{2ik\eta_{2}}}{2\kappa^{2}}C_{1}+z_{\kappa}^{*}C_{2}\thinspace,\label{eq:-30}
\end{align}
where $\kappa\equiv k(\eta_{2}-2\eta_{1})$ and $z_{\kappa}\equiv1-\frac{i}{\kappa}-\frac{1}{2\kappa^{2}}$. 

The resulting phase space distribution function reads:
\begin{align}
f(k)= & \frac{1}{4k^{4}\Delta_{2}^{4}}+\frac{1}{k^{6}\eta_{1}^{6}}+\frac{1}{2k^{10}\eta_{1}^{6}\Delta_{2}^{4}}\nonumber \\
 & -\frac{2k^{4}\eta_{1}^{2}\Delta_{2}\left(\Delta_{1}+\Delta_{2}\right)-2k^{2}\Delta_{1}^{2}+1}{2k^{10}\eta_{1}^{6}\Delta_{2}^{4}}\cos\left(2k\Delta_{1}\right)\nonumber \\
 & -\frac{2k^{5}\eta_{1}^{3}\Delta_{2}^{2}-k^{3}\eta_{1}\left(\Delta_{1}+\Delta_{2}\right){}^{2}+2k\Delta_{1}}{2k^{10}\eta_{1}^{6}\Delta_{2}^{4}}\sin\left(2k\Delta_{1}\right),\label{eq:-27}
\end{align}
where $\Delta_{1}=\eta_{2}-\eta_{1}$ and $\Delta_{2}=\eta_{2}-2\eta_{1}$.
 If we take the limit $\eta_{2}\to\eta_{1}$ (equivalent to $\Delta_{2}^{2}\to\eta_{1}^{2}$),
it is straightforward to verify that Eq.~\eqref{eq:-27} reduces to
Eq.~\eqref{eq:-9-1}. 

\begin{figure}
\centering

\includegraphics[width=0.495\textwidth]{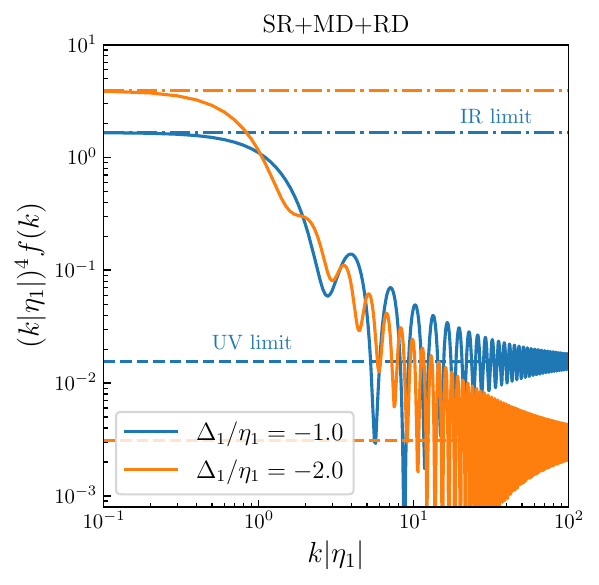}\includegraphics[width=0.495\textwidth]{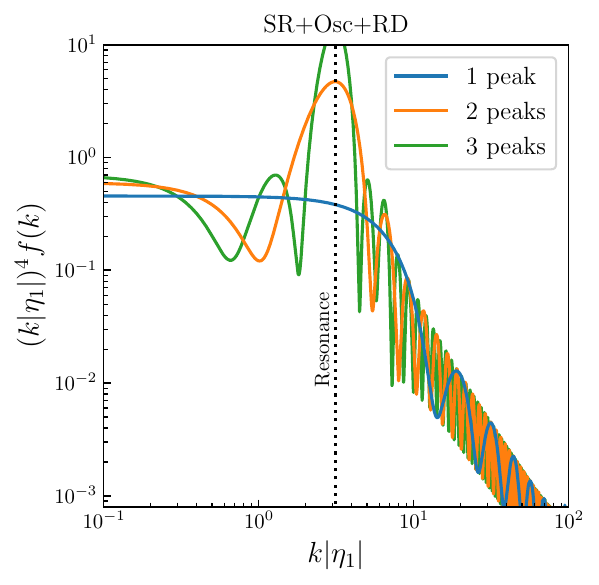}

\caption{\emph{Left Panel}: Graviton production from the spacetime background
that evolves from SR to MD and RD in the analytical form given by
Eq.~\eqref{eq:-28}. The solid curves are produced using Eq.~\eqref{eq:-27}.
 The dashed and dash-dotted lines represent the UV and IR limits
given by Eqs.~\eqref{eq:-32} and \eqref{eq:-33}, respectively. \emph{Right
Panel}: Similar to the left panel but with the MD epoch replaced by
an oscillatory one, as formulated in Eq.~\eqref{eq:-38}. \label{fig:ana-sol}}

\end{figure}

In the left panel of Fig.~\ref{fig:ana-sol}, we plot $k^{4}|\eta_{1}|^{4}f(k)$
with $f$ given in Eq.~\eqref{eq:-27} for two samples, $\Delta_{1}/|\eta_{1}|=1$
and $2$. The corresponding values of $\Delta_{2}$ are $2|\eta_{1}|$
and $3|\eta_{1}|$, respectively. The curves are flat at small $k$,
highly oscillatory at medium $k$, and eventually converging to certain
nonzero values. Both the low-$k$ (IR) and high-$k$ (UV) limits can
be derived analytically. If $k$ is sufficiently large, one can expand
Eq.~\eqref{eq:-27} in terms of $k^{-1}$ and keep only the leading
order: 
\begin{equation}
f(k)\approx\frac{1}{4k^{4}\Delta_{2}^{4}}+{\cal O}\left(k^{-5}\right).\label{eq:-32}
\end{equation}
If $k$ is sufficiently small, one can expand it in terms of $k$
and obtain
\begin{equation}
f(k)\approx\frac{1}{36k^{4}\Delta_{2}^{4}}\left[\left(4\frac{\Delta_{2}^{3}}{\eta_{1}^{3}}+1\right)^{2}+{\cal O}\left(k^{2}\right)\right].\label{eq:-33}
\end{equation}

It is interesting to notice that, while $f$ in Eq.~\eqref{eq:-27}
contains terms proportional to $k^{-10}$ and $k^{-6}$ which could
dominate over $k^{-4}$ in the IR limit, these terms cancel out in
the series expansion, leading to $f\propto k^{-4}\left[1+{\cal O}\left(k^{2}\right)\right]$,
which guarantees that the IR limit of $\Omega_{{\rm GW}}$ is still
scale-invariant.

More interestingly, if $\Delta_{2}\gg|\eta_{1}|$, the IR limit of
$f(k)$ in Eq.~\eqref{eq:-33} can be substantially enhanced by $\Delta_{2}/|\eta_{1}|$
but due to a cancellation between this enhancement and the expansion
factor in Eq.~\eqref{eq:-34}, the resulting $\Omega_{{\rm GW}}$
only exhibits weak dependence on $\Delta_{2}/|\eta_{1}|$. Substituting
Eq.~\eqref{eq:-33} into Eq.~\eqref{eq:-17} and using Eq.~\eqref{eq:-34},
we obtain
\begin{equation}
\Omega_{{\rm GW}}\simeq2.3\times10^{-16}\left(\frac{H_{I}}{10^{14}\ \text{GeV}}\right)^{2}\left(\frac{g_{\star}}{106.75}\right)^{-\frac{1}{3}}\left[\frac{4}{3}-\frac{1}{3}\left(\frac{a_{e}}{a_{{\rm rh}}}\right)^{\frac{3}{2}}\right]^{2},\label{eq:-26-1}
\end{equation}
which differs from Eq.~\eqref{eq:-26} by an ${\cal O}(1)$ factor.
For $a_{{\rm rh}}/a_{e}$ varying from $1$ to $\infty$, this factor
varies from $1$ to $16/9$. 

Again, here we see a similar UV divergence appearing in $\rho_{{\rm GW}}\propto\int^{\infty}fk^{3}dk$.
However, the UV divergence here can be suppressed, in principle, by
increasing the length of the MD epoch (corresponding to increasing
$\Delta_{1}$ or $\Delta_{2}$). As is shown in the left panel of
Fig.~\ref{fig:ana-sol}, the orange line has a lower UV limit than
the blue line. Nevertheless, as long as there is a sharp cut-off in
Eq.~\eqref{eq:-28} at $\eta_{2}$, the UV divergence always exists,
as already discussed in Sec.~\ref{subsec:SR+RD}.

\subsection{SR+KD+RD}

Since the inflaton energy density and pressure are determined by $\rho_{\phi}=\dot{\phi}^{2}/2+V$
and $p_{\phi}=\dot{\phi}^{2}/2-V$ with $\phi$ the inflaton field
and $V$ the inflaton potential, it is possible that $\rho_{\phi}\approx p_{\phi}$
may occur at the end of inflation. This is known as kination domination
(KD), during which $\mu^{2}$ scales as $-\frac{1}{4\eta^{2}}$. 
Motivated by this scenario, we construct the following $\mu^{2}$
function:
\begin{equation}
\mu^{2}\left(\eta\right)=\begin{cases}
\frac{2}{\eta^{2}} & \eta\leq\eta_{1}\\
-\frac{1}{4\eta^{2}}+\frac{9}{4\eta_{1}^{2}} & \eta_{1}<\eta\leq\frac{\eta_{1}}{3}\\
0 & \frac{\eta_{1}}{3}<\eta
\end{cases}\thinspace,\label{eq:-31}
\end{equation}
with $\eta_{1}<0$.  We have added $9/4\eta_{1}^{2}$ in the second
line to ensure that $\mu^{2}$ is continuous at $\eta=\eta_{1}$.
The second transitioning point is set at $\eta=\eta_{1}/3$ also for
continuity. 

With $\mu^{2}$ given in Eq.~\eqref{eq:-31}, we solve $\chi_{k}$
as follows:
\begin{equation}
\chi_{k}(\eta)=\begin{cases}
\frac{1}{\sqrt{2k}}\left(1-\frac{i}{k\eta}\right)e^{-ik\eta} & \eta\le\eta_{1}\\
C_{1}\sqrt{\eta}J_{0}(\omega_{k}^{{\rm II}}\eta)+C_{2}\sqrt{\eta}Y_{0}(\omega_{k}^{{\rm II}}\eta) & \eta_{1}<\eta\le\frac{\eta_{1}}{3}\\
\alpha\frac{e^{-ik\eta}}{\sqrt{2k}}+\beta\frac{e^{ik\eta}}{\sqrt{2k}} & \frac{\eta_{1}}{3}<\eta
\end{cases},\label{eq:-57}
\end{equation}
where $\omega_{k}^{{\rm II}}\equiv\sqrt{k^{2}-9/(4\eta_{1}^{2})}$,
and $J$ and $Y$ are the Bessel functions of the first and second
kinds. Imposing continuity conditions on $\chi_{k}$ and $\chi'_{k}$
at $\eta=\eta_{1}$ and $\eta_{1}/3$, we obtain
\begin{align}
C_{1} & =\frac{e^{-ik\eta_{1}}}{2\sqrt{2}i}\cdot\frac{-z_{2}Y_{0}(x_{1})+2x_{1}(1+ik\eta_{1})Y_{1}(x_{1})}{(k\eta_{1})^{3/2}x_{1}\left[J_{0}(x_{1})Y_{1}(x_{1})-J_{1}(x_{1})Y_{0}(x_{1})\right]}\thinspace,\label{eq:-58}\\
C_{2} & =\frac{e^{-ik\eta_{1}}}{2\sqrt{2}i}\cdot\frac{z_{2}J_{0}(x_{1})-2x_{1}(1+ik\eta_{1})J_{1}(x_{1})}{(k\eta_{1})^{3/2}x_{1}\left[J_{0}(x_{1})Y_{1}(x_{1})-J_{1}(x_{1})Y_{0}(x_{1})\right]}\thinspace,\label{eq:-59}\\
\alpha & =\frac{e^{ik\eta_{1}/3}}{2}\cdot\frac{C_{1}\left[z_{1}J_{0}\left(\frac{x_{1}}{3}\right)-2ix_{1}J_{1}\left(\frac{x_{1}}{3}\right)\right]+C_{2}\left[z_{1}Y_{0}\left(\frac{x_{1}}{3}\right)-2ix_{1}Y_{1}\left(\frac{x_{1}}{3}\right)\right]}{\sqrt{6k\eta_{1}}}\thinspace,\label{eq:-60}\\
\beta & =\frac{e^{-ik\eta_{1}/3}}{2}\cdot\frac{C_{1}\left[z_{1}^{*}J_{0}\left(\frac{x_{1}}{3}\right)+2ix_{1}J_{1}\left(\frac{x_{1}}{3}\right)\right]+C_{2}\left[z_{1}^{*}Y_{0}\left(\frac{x_{1}}{3}\right)+2ix_{1}Y_{1}\left(\frac{x_{1}}{3}\right)\right]}{\sqrt{6k\eta_{1}}}\thinspace,\label{eq:-61}
\end{align}
where $x_{1}\equiv\omega_{k}^{{\rm II}}\eta_{1}$, $z_{1}\equiv3i+2k\eta_{1}$
and $z_{2}\equiv3+3ik\eta_{1}-2k^{2}\eta_{1}^{2}$. 

The resulting $f(k)$ is quite lengthy so we do not present it here.
But it is straightforward to construct it using the above expressions.
Here we would like to inspect its UV limit, which can be obtained
by expanding the above results in $(k\eta_{1})^{-1}$ and computing
the corresponding $f(k)$. The result reads
\begin{equation}
\lim_{k\to\infty}f(k)=\frac{3^{4}}{2^{18}(k\eta_{1})^{6}}\left|23i-96ie^{4ik\eta_{1}/3}+9ie^{2ik\eta_{1}}\right|^{2}.\label{eq:-62}
\end{equation}
At large $k$, this result decreases faster than $k^{-4}$, implying
that there is no UV divergence in the total energy density of GWs.
This example demonstrates that the continuity of $\mu^{2}$ can indeed
lead to $\underset{k\to\infty}{{\rm lim}}fk^{4}\to0$  so as to avoid
UV divergences.

\subsection{SR+Oscillations \label{subsec:SR+Osc}}

Finally, let us consider an oscillatory scenario that allows for analytical
solutions:
\begin{equation}
\mu^{2}\left(\eta\right)=\begin{cases}
\frac{2}{\eta^{2}} & \eta\leq\eta_{1}\\
\frac{2}{\eta_{1}^{2}}\cos\left(2\pi\eta/\eta_{1}\right) & \eta_{1}<\eta\leq\eta_{2}\\
0 & \eta_{2}<\eta
\end{cases}\thinspace,\label{eq:-38}
\end{equation}
where $\eta_{2}=\eta_{1}+|\eta_{1}|(n-3/4)$ with $n\in\{1,2,3,\cdots\}$.
The $\mu^{2}$ function constructed in this way is always continuous
at $\eta=\eta_{1}$ and $\eta_{2}$. The integer $n$ can be identified
as the number of peaks of the $\mu^{2}$ curve. 

With $\mu^{2}$ given by Eq.~\eqref{eq:-38}, we obtain the following
solution
\begin{equation}
\chi_{k}(\eta)=\begin{cases}
\frac{1}{\sqrt{2k}}\left(1-\frac{i}{k\eta}\right)e^{-ik\eta} & \eta\leq\eta_{1}\thinspace,\\
\frac{C_{1}}{\sqrt{2k}}\mathds{C}\left(\frac{k^{2}\eta_{1}^{2}}{\pi^{2}},\frac{1}{\pi^{2}},\frac{\pi\eta}{\eta_{1}}\right)+\frac{C_{2}}{\sqrt{2k}}\mathds{S}\left(\frac{k^{2}\eta_{1}^{2}}{\pi^{2}},\frac{1}{\pi^{2}},\frac{\pi\eta}{\eta_{1}}\right) & \eta_{1}<\eta\leq\eta_{2}\\
\frac{\alpha}{\sqrt{2k}}e^{-ik\eta}+\frac{\beta}{\sqrt{2k}}e^{ik\eta} & \eta_{2}<\eta\thinspace,
\end{cases}\thinspace,\label{eq:-39}
\end{equation}
with 
\begin{align}
C_{1} & =\frac{is_{1}\left(1+ik\eta_{1}-k^{2}\eta_{1}^{2}\right)+\pi(i-k\eta_{1})s'_{1}}{\pi k\eta_{1}\left(c'_{1}s_{1}-c_{1}s'_{1}\right)}e^{-ik\eta_{1}}\thinspace,\label{eq:-40}\\
C_{2} & =\frac{ic_{1}\left(1+ik\eta_{1}-k^{2}\eta_{1}^{2}\right)+\pi(i-k\eta_{1})c'_{1}}{\pi k\eta_{1}\left(c_{1}s'_{1}-c'_{1}s_{1}\right)}e^{-ik\eta_{1}}\thinspace,\label{eq:-41}\\
\alpha & =\frac{C_{1}\left(ik\eta_{1}c_{2}-\pi c'_{2}\right)+C_{2}\left(ik\eta_{1}s_{2}-\pi s'_{2}\right)}{2ik\eta_{1}}e^{ik\eta_{2}}\thinspace,\label{eq:-42}\\
\beta & =\frac{C_{1}\left(ik\eta_{1}c_{2}+\pi c'_{2}\right)+C_{2}\left(ik\eta_{1}s_{2}+\pi s'_{2}\right)}{2ik\eta_{1}}e^{-ik\eta_{2}}\thinspace,\label{eq:-43}
\end{align}
where $c_{i}\equiv\mathds{C}\left(\frac{k^{2}\eta_{1}^{2}}{\pi^{2}},\frac{1}{\pi^{2}},\pi\frac{\eta_{i}}{\eta_{1}}\right)$
and $s_{i}\equiv\mathds{S}\left(\frac{k^{2}\eta_{1}^{2}}{\pi^{2}},\frac{1}{\pi^{2}},\pi\frac{\eta_{i}}{\eta_{1}}\right)$
with $i=1$ or $2$, and $c'_{i}$ ($s'_{i}$) denotes the derivative
with respect to the last argument of the $\mathds{C}$ ($\mathds{S}$)
function. 

The resulting expression of $f$ is also too lengthy to be presented
here while its calculation from  the above expressions is straightforward.
In the right panel of Fig.~\ref{fig:ana-sol}, we plot $k^{4}|\eta_{1}|^{4}f(k)$
for this oscillatory scenario with $n=1$, $2$, and 3 (the number
of peaks of the $\mu^{2}$ curve is equal to $n$). Similar to the
left panel, here we see that these curves are also flat at small $k$
and highly oscillatory at medium $k$. The most prominent differences
are that (i) they eventually converge to zero as a consequence of
continuous $\mu^{2}$ and (ii) the orange ($n=2$) and green $(n=3)$
curves exhibit significant resonances at $k|\eta_{1}|\approx\pi$.
This can be understood as resonance production: when $k$ matches
the frequency of the oscillation in Eq.~\eqref{eq:-38}, the corresponding
mode function is enhanced efficiently by the background oscillations. 

\section{The $D$ parametrization and UV smoothing\label{sec:D-parametrization}}

\begin{figure}
\centering

\includegraphics[width=0.9\textwidth]{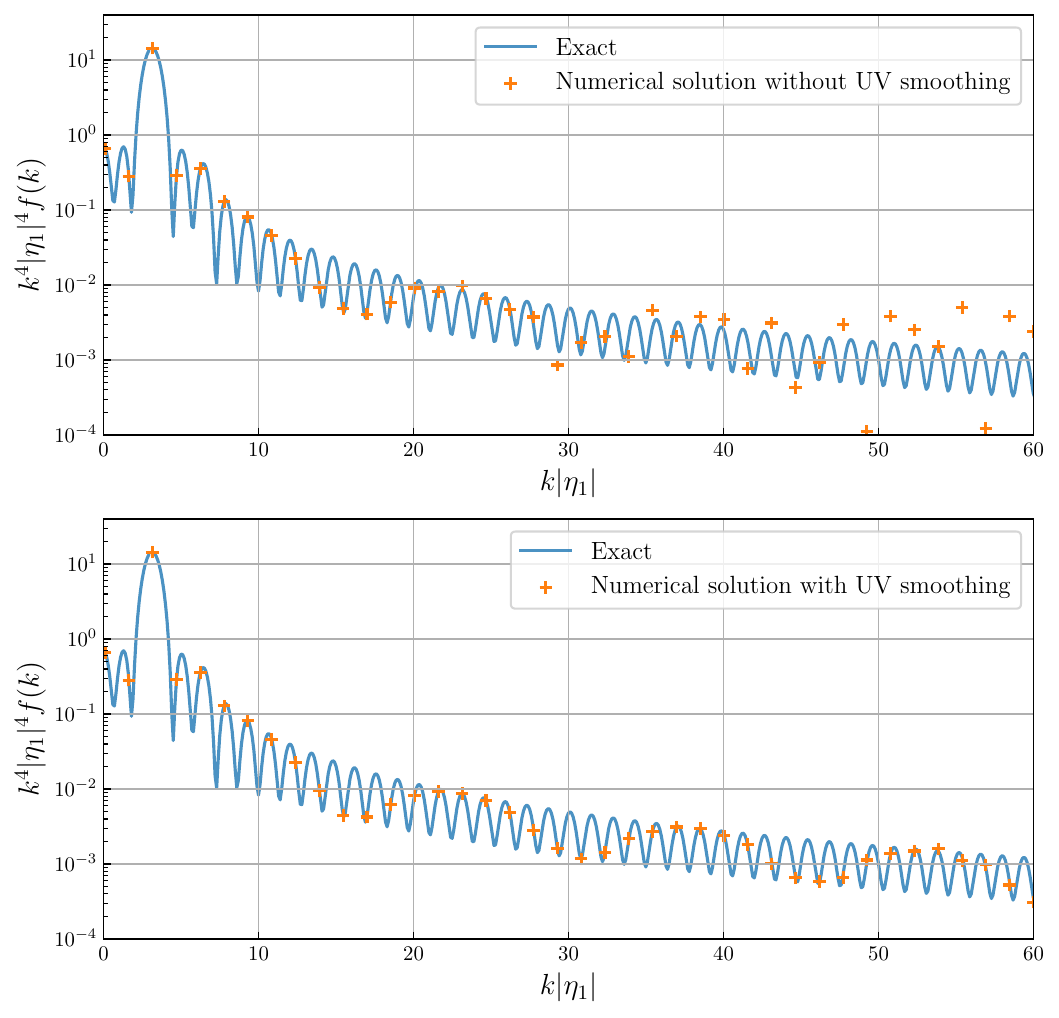}\caption{\label{fig:UV-smooth}Numerical results (orange points) before and
after applying the UV smoothing technique, compared with exact ones
(blue curves). Here we assume that $\mu^{2}$ takes the form in Eq.~\eqref{eq:-38},
for which exact analytical results can be obtained.}
\end{figure}

 From the analytical studies in Sec.~\ref{sec:Analytical-solutions},
we have seen two issues that could be troublesome for numerical calculations.
First, large cancellations could be hidden in  Eq.~\eqref{eq:-3},
rendering a straightforward calculation based on the original mode
equation vulnerable to numerical inaccuracies. Second, any discontinuities
in the $\mu^{2}$ function, no matter how small, would lead to UV
divergences in the total energy of produced GWs. In this section,
we present our numerical techniques to address these issues. The first
issue is resolved by a new parametrization, which is similar to the
widely used $\alpha$-$\beta$ parametrization in Eq.~\eqref{eq:-7}
but remains valid in the presence of the tachyonic mode issue. The
second issue is resolved by UV smoothing, which involves the adiabaticity
parameter to control the smoothness of $\mu^{2}$ and can significantly
improve the numerical calculation (see Fig.~\ref{fig:UV-smooth}). 

The new parametrization we adopt is 
\begin{equation}
\chi(\eta)=\frac{1}{\sqrt{2k}}\left[1+D(\eta)\right]e^{-ik\eta}\thinspace,\label{eq:-12}
\end{equation}
where $D$ quantifies the deviation of $\chi(\eta)$ from the plane-wave
solution. Hence we refer to it as the $D$ parametrization. 

Using Eq.~\eqref{eq:-12}, Eq.~\eqref{eq:} can be rewritten as
\begin{equation}
D''(\eta)-2ikD'(\eta)-\left[1+D(\eta)\right]\mu^{2}(\eta)=0\thinspace,\label{eq:-13}
\end{equation}
with the initial condition 
\begin{equation}
\lim_{\eta\to-\infty}D(\eta)=\lim_{\eta\to-\infty}D'(\eta)=0\thinspace.\label{eq:-44}
\end{equation}

The Wronskian condition in terms of $D$ reads
\begin{equation}
\frac{\left|D\right|^{2}}{2}+\text{Re}D-{\rm Im}\left[\frac{(1+D^{*})D'}{2k}\right]=0\thinspace.\label{eq:-15}
\end{equation}
Using the above Wronskian condition, we find that Eq.~\eqref{eq:-3}
simplifies to
\begin{equation}
f=\lim_{\eta\to\infty}\frac{\left|D'\right|^{2}}{4k^{2}}\thinspace.\label{eq:-14}
\end{equation}
An advantage of Eq.~\eqref{eq:-14} is that, unlike Eq.~\eqref{eq:-3},
it does not contain terms much greater than the resulting $f$, thereby
avoiding potentially large cancellations similar to what we have seen
in Eqs.~\eqref{eq:-35} and \eqref{eq:-36}.  

Ideally, one should solve the differential equation from $\eta=-\infty$
to a sufficiently late epoch with vanishing $\mu^{2}$ (e.g., deep
into the RD epoch). In practice, the initial point has to be set at
a finite $\eta$ (denoted by $\eta_{i}$ below). Besides, for very
large $k$, an excessively long period of evolution is computationally
expensive and actually unnecessary. Therefore, it is important to
identify the relevant time window for graviton production. 

In general, when the following condition is satisfied, the corresponding
time window is considered irrelevant:
\begin{equation}
k^{2}\gg\left|\mu^{2}\right|\ \land\ \left|A_{k}\right|\ll\frac{\left|\mu^{2}\right|}{k^{2}}\thinspace,\label{eq:-45}
\end{equation}
where $A_{k}$ is the adiabaticity parameter, defined as
\begin{equation}
A_{k}\equiv\frac{\omega_{k}'}{\omega_{k}^{2}}\thinspace.\label{eq:-46}
\end{equation}
Note that the second condition in Eq.~\eqref{eq:-45} is stronger
than the usual adiabaticity condition $A_{k}\ll1$, which only guarantees
that non-adiabatic effects on $\chi_{k}$ are weak. The stronger adiabaticity
condition $\left|A_{k}\right|\ll\left|\mu^{2}\right|/k^{2}$ ensures
that non-adiabatic effects on $k^{4}f$ are suppressed (see Appendix~\ref{sec:Suppression-of-UV}
for a concrete example).  When $k^{2}\gg\left|\mu^{2}\right|$ is
satisfied, $\left|A_{k}\right|\ll\left|\mu^{2}\right|/k^{2}$ is equivalent
to $\left|d\ln\mu^{2}/d\eta\right|/k\ll1$. 

In the SR epoch where $\mu^{2}\approx2/\eta^{2}$, if $k^{2}\gg\left|\mu^{2}\right|$
is satisfied (corresponding to $|k\eta|^{2}\text{\ensuremath{\gg}}2$),
then $\left|A_{k}\right|\approx2/|k\eta|^{3}\ll\left|\mu^{2}\right|/k^{2}$,
implying that the second requirement in Eq.~\eqref{eq:-45} is also
satisfied. However, if one sets the initial point in the SR epoch
with Eq.~\eqref{eq:-45} well satisfied,  there is still a subtle
issue that could lead to numerical errors---see the upper panel of
Fig.~\ref{fig:UV-smooth}, where we set the initial point at $\eta_{i}=-200/k$
and try to solve the differential equation numerically with the analytical
$\mu^{2}$ function in Eq.~\eqref{eq:-38}. The blue line represents
the exact result obtained from the analytical calculation in Sec.~\ref{subsec:SR+Osc},
and the orange points are numerical results. Obviously the numerical
results are inaccurate or even erroneous at large $k$. 

This numerical inaccuracy  is caused by a small mismatch of the initial
values of $D(\eta_{i})$ and $D'(\eta_{i})$ with their true values.
When numerically solving the differential equation starting at $\eta_{i}$,
if $D(\eta_{i})=D'(\eta_{i})=0$ is used, it is technically equivalent
to solving the equation starting from $\eta=-\infty$ with $\mu^{2}(\eta)\to\mu^{2}(\eta)\,\Theta(\eta-\eta_{i})$.
The two differential-equation systems share the same solution at $\eta\geq\eta_{i}$.
Therefore, we can employ the latter to understand the behavior of
the former. The deviation of $D(\eta_{i})$ and $D'(\eta_{i})$ from
their true values in the first system is related to the abrupt change
of $\mu^{2}$ at $\eta=\eta_{i}$ in the second system, in which a
plane wave of the initial form $e^{-ik\eta}/\sqrt{2k}$ scatters towards
the potential $\mu^{2}$ that only appears at $\eta\geq\eta_{i}$.
Consequently, UV noise for large $k$ modes is expected to appear
in the second system, and by equivalence, also in the first system.
 One could consider improving the accuracy of the initial conditions
using analytical results in the SR epoch. However, in a concrete inflationary
model, the background is not exactly de Sitter space, implying that
the SR approximation can still lead to small inaccuracies in the initial
condition. One could also push $\eta_{i}$ earlier such that the Bunch-Davies
initial conditions become more accurate. For instance, setting $\eta_{i}=-1600/k$
instead of $\eta_{i}=-200/k$ in the upper panel of Fig.~\ref{fig:UV-smooth}
would lead to sufficient accuracy within the shown range of $k$.
But for larger $k$, the same problem would appear again and the numerical
solutions become significantly more computationally expensive. 

A more practical method is to suppress $\mu^{2}$ around $\eta_{i}$
such that $\mu^{2}(\eta_{i})=0$ and let it smoothly transition from
zero to physical values of $\mu^{2}$.  This approach is effective
as long as the transition completes within the irrelevant time window
defined by Eq.~\eqref{eq:-45}, plus an extra requirement that the
transition also satisfies Eq.~\eqref{eq:-45}. We refer to this procedure
as {\it UV smoothing}. Technically, this is achieved by multiplying
an exponential factor to $\mu^{2}$ as follows
\begin{equation}
\mu^{2}\left(\eta\right)\to\exp\left[-\epsilon\left(\frac{\eta_{b}-\eta_{a}}{\eta-\eta_{a}}\right)^{2}\right]\Theta\left(\eta-\eta_{a}\right)\mu^{2}\left(\eta\right),\label{eq:-47}
\end{equation}
 which contains three free parameters, $\epsilon$, $\eta_{a}$,
and $\eta_{b}$. Here  $\epsilon$ is positive, and $\eta_{a}$ and
$\eta_{b}$ satisfy $\eta_{i}\leq\eta_{a}<\eta_{b}$. For $\eta<\eta_{a}$,
Eq.~\eqref{eq:-47} sets $\mu^{2}$ to zero. Starting from $\eta_{a}$,
it gradually deviates from zero. Despite the Heaviside theta function
in Eq.~\eqref{eq:-47}, it is smooth at $\eta=\eta_{a}$ (mathematically,
it is infinitely differentiable at this point). When $\eta$ increases
to $\eta_{b}$, Eq.~\eqref{eq:-47} gives $\mu^{2}\left(\eta\right)\to e^{-\epsilon}\mu^{2}\left(\eta\right)$,
implying that $\mu^{2}$ approximately recovers its original value
at $\eta=\eta_{b}$. 

In the lower panel of Fig.~\ref{fig:UV-smooth}, we demonstrate that
the UV smoothing technique can significantly improve the numerical
calculation. In this plot, we set $\eta_{i}=-200/k$ (same as the
upper panel), and use Eq.~\eqref{eq:-47} with $\epsilon=0.1$, $\eta_{a}=0.98\eta_{i}$,
$\eta_{b}=0.9\eta_{i}$. The numerical results are in excellent agreement
with the exact ones. 

The analytical expression in Eq.~\eqref{eq:-47} can be used to derive
a criterion for selecting the parameters $\epsilon$, $\eta_{a}$,
and $\eta_{b}$ in the SR regime. Through the calculation in Appendix~\ref{sec:adiabatic},
we find that the adiabaticity condition $\left|A_{k}\right|\ll\left|\mu^{2}\right|/k^{2}$
approximately corresponds to
\begin{equation}
\sqrt{\frac{27}{8\epsilon}}\frac{e^{-3/2}}{k(\eta_{b}-\eta_{a})}\ll1\thinspace.\label{eq:-66}
\end{equation}
For  $\epsilon=0.1$, $\eta_{a}=0.98\eta_{i}$, and $\eta_{b}=0.9\eta_{i}$
used in the above example, the left-hand side of Eq.~\eqref{eq:-66}
is $0.08\ll1$. Hence the adiabaticity condition is well satisfied.
More generally, one can verify that for $\epsilon$, $\eta_{a}$,
and $\eta_{b}$ satisfying Eq.~\eqref{eq:-66} and $k|\eta_{i}|>k|\eta_{a}|>k|\eta_{b}|\gg1$,
the impact of varying these parameters on the final result is negligibly
small. 

Let us finally comment on the necessity of UV smoothing in this work.
In principle, pushing $\eta_{i}$ to a sufficiently early point should
be equivalent to the UV smoothing technique applied here. In practice,
the former would require tracking the evolution of the mode function
from a highly oscillatory phase deep in the sub-horizon regime. The
computational cost of tracking such an oscillatory phase is unnecessarily
high, while the most computationally expensive sub-horizon regime
is not the physical regime of interest. Our UV smoothing method avoids
this overhead, ensuring that the computational cost is effectively
allocated to the physically relevant epoch.

\section{GW spectra in specific inflationary models \label{sec:Result}}

In this section, we apply the numerical framework developed in Sec.~\ref{sec:D-parametrization}
to compute the primordial GW spectra for two  phenomenologically
viable inflationary models:  the $\alpha$-attractor T model and
the Starobinsky model. Although the two models lead to identical predictions
on the tensor-to-scalar ratio, $r\approx12/N_{e}^{2}$ with $N_{e}=50\sim60$
the number of e-folds, we will show that due to its less harmonic
behavior around the minimum, the Starobinsky model leads to a more
oscillatory GW spectrum at high frequencies than the T model.

We assume that reheating is achieved by a constant decay rate of inflatons
to radiation, denoted by $\Gamma_{\phi}$. Then the evolution of the
inflaton field $\phi$ and the radiation energy density $\rho_{r}$
is governed by 
\begin{align}
\ddot{\phi}+(3H+\Gamma_{\phi})\dot{\phi} & =-\frac{dV}{d\phi}\thinspace,\label{eq:-51}\\
\dot{\rho}_{r}+4H\rho_{r} & =\Gamma_{\phi}\left(\rho_{\phi}+p_{\phi}\right),\label{eq:-52}
\end{align}
where $\rho_{\phi}=\frac{1}{2}\dot{\phi}^{2}+V$ and $p_{\phi}=\frac{1}{2}\dot{\phi}^{2}-V$,
with $V$ the inflaton potential. The Hubble parameter is determined
by
\begin{equation}
H=\sqrt{\frac{\rho_{\phi}+\rho_{r}}{3M_{P}^{2}}}\thinspace.\label{eq:-53}
\end{equation}
Given a specific form of $V$, it is straightforward to solve Eqs.~\eqref{eq:-51}
and \eqref{eq:-52}. With the obtained results of $p_{\phi}$ and
$\rho_{\phi}$, we use Eq.~\eqref{eq:-50} to compute the Ricci scalar
$R$ and $\mu^{2}=-a^{2}R/6$. Radiation does not contribute to the
Ricci scalar according to the last expression in Eq.~\eqref{eq:-50}. 

Note that the solution of $\phi$ and the resulting $R$ and $\mu^{2}$
at late times can be highly oscillatory---see Fig.~\ref{fig:mu2-T}
for illustration. Such oscillations are responsible for the production
of high-frequency GWs so they need to be evaluated accurately. To
this end, we impose a maximal step size in the differential equation
solver. In {\tt scipy.integrate.solve\_ivp}, this is achieved by
setting {\tt max\_step=}$\Delta t/N$ where $\Delta t=2\pi/m_{\phi}$
is the oscillation period and $N$ a relatively large number to ensure
the accuracy. For a very large number of oscillations, numerical evaluation
of these background quantities ($\phi$, $R$, and $\mu^{2}$) can
be computationally expensive. In this case, one can use the analytical
solutions presented in Appendix B of Ref.~\cite{Mojahed:2024mvb}
to extrapolate the numerical results.

\subsection{The $\alpha$-attractor T model \label{subsec:T}}

The $\alpha$-attractor T model~\cite{Kallosh:2013hoa} has the following
potential:
\begin{equation}
V=\lambda M_{P}^{4}\left[\sqrt{6}\tanh\left(\frac{\phi}{\sqrt{6}M_{P}}\right)\right]^{2n},\label{eq:-54}
\end{equation}
with $n=1$, $2$, $3$, $\cdots$. The most often considered case
is $n=1$, which leads to harmonic oscillations around the minimum
of the potential: $V\approx\frac{m_{\phi}^{2}}{2}\phi^{2}$ with $m_{\phi}^{2}\equiv2\lambda M_{P}^{2}$.
 Here the constant $\lambda$ is determined by $\lambda\approx3\pi^{2}A_{s}/N_{e}^{2}\approx2.055\times10^{-11}$~\cite{Barman:2022qgt},
where $A_{s}\approx2.1\times10^{-9}$~\cite{Planck:2018jri} is the
amplitude of scalar fluctuations and $N_{e}\approx55$ is the number
of e-folds. Correspondingly, the mass is determined to be $m_{\phi}\approx6.41\times10^{-6}M_{P}$.
In our analysis for the T model, we focus on the case of $n=1$. 

With Eq.~\eqref{eq:-54}, we solve Eqs.~\eqref{eq:-51} and \eqref{eq:-52}
to obtain the $\mu^{2}(\eta)$ function, which is then used in Eq.~\eqref{eq:-13}
to compute the mode function. 

\begin{figure}
\centering

\includegraphics[width=0.8\textwidth]{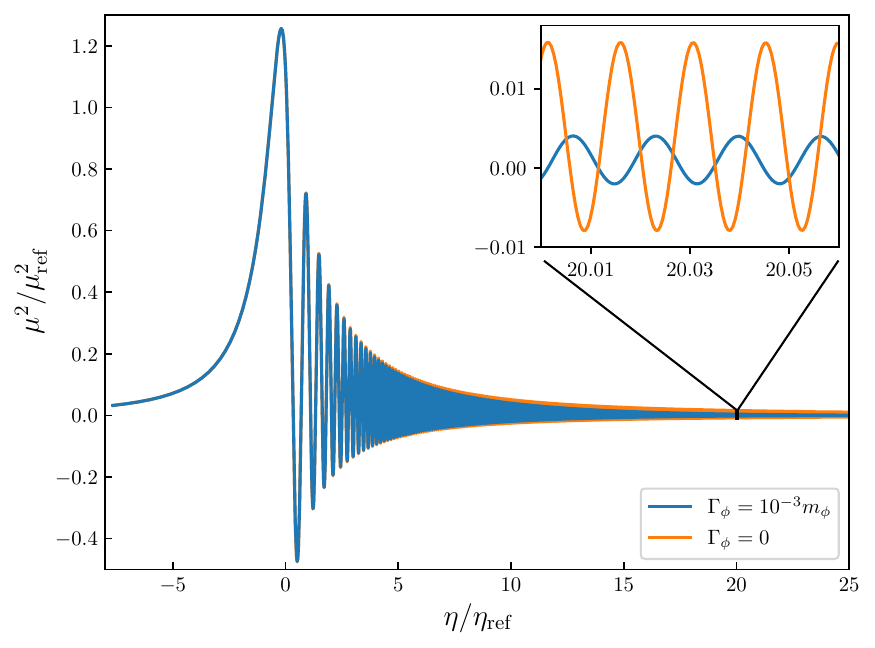}\caption{\label{fig:mu2-T} The $\mu^{2}(\eta)$ function in the T model with
$\Gamma_{\phi}=10^{-3}m_{\phi}$ (blue) and $\Gamma_{\phi}=0$ (orange). }
\end{figure}

\begin{figure}
\centering

\includegraphics[height=0.45\textwidth]{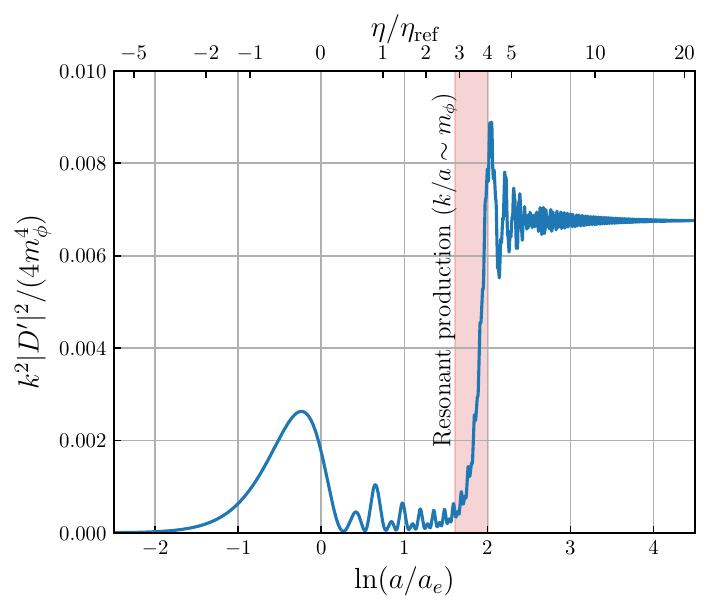}\includegraphics[height=0.45\textwidth]{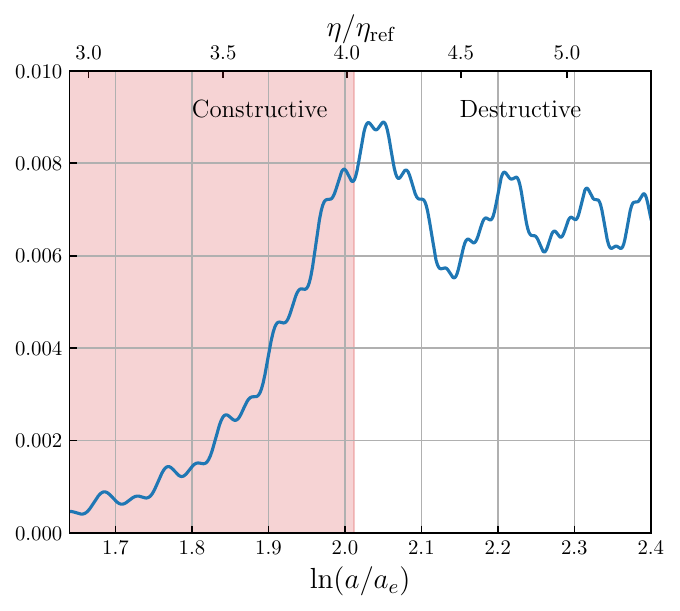}\caption{\label{fig:fk4}\emph{ Left Panel}: The evolution curve of $k^{2}\left|D'\right|^{2}/(4m_{\phi}^{4})$,
whose final value is proportional to $\Omega_{\text{GW}}h^{2}$. This
curve shows how high-frequency GWs in the T model are produced from
the oscillatory background---when its physical momentum matches the
frequency of the background oscillation ($k/a\sim m_{\phi}$), resonant
production is attained, leading to efficient production of gravitons
in the light red region. The plot is produced with $k=15.4k_{{\rm ref}}$
and $\Gamma_{\phi}=10^{-3}m_{\phi}$.  \emph{Right Panel}: Similar
to the left but focusing on a shorter period  to demonstrate constructive
and destructive interference among the large number of oscillations. }
\end{figure}

In Fig.~\ref{fig:mu2-T}, we plot the $\mu^{2}(\eta)$ function for
two cases, $\Gamma_{\phi}=10^{-3}m_{\phi}$ and $\Gamma_{\phi}=0$.
The x-axis and y-axis are normalized by $\eta_{{\rm ref}}\equiv1/(aH)_{{\rm max}}$
and $\mu_{{\rm ref}}^{2}\equiv(aH)_{{\rm max}}^{2}$, respectively.
Here $(aH)_{{\rm max}}$ is the maximal value of $aH$, which increases
during SR and decreases during MD and RD. Hence $(aH)_{{\rm max}}$
can be approximately viewed as the value of $aH$ at the end of inflation.
As is shown in this figure, the $\mu^{2}(\eta)$ function is highly
oscillatory at late times. These oscillations will eventually be suppressed
by a finite decay rate of inflatons. For instance, the blue curve
at $\eta\approx20\eta_{{\rm ref}}$ oscillates with significantly
lower amplitudes than the orange curve.  

In Fig.~\ref{fig:fk4}, we show the evolution curve of $k^{2}\left|D'\right|^{2}/(4m_{\phi}^{4})$
with $k=15.4k_{{\rm ref}}$ where $k_{{\rm ref}}\equiv(aH)_{{\rm max}}$.
According to Eq.~\eqref{eq:-14}, this quantity is proportional to
$fk^{4}\propto\Omega_{\text{GW}}h^{2}$ so it can be used to illustrate
how high-frequency GWs are produced during the evolution. The bottom
x-axis represents $\ln(a/a_{e})$ with $a_{e}$ the scale factor at
$\eta=0$, and the top x-axis indicates the corresponding values of
$\eta/\eta_{{\rm ref}}$. In the left panel, we see that the widest
peak occurs in $-5\lesssim\eta/\eta_{{\rm ref}}\lesssim0$, corresponding
to the first peak in Fig.~\ref{fig:mu2-T}. However, due to the high
frequency of the mode function ($k\gg k_{{\rm ref}}$), it increases
and decreases adiabatically. Consequently, each oscillation of $\mu^{2}(\eta)$
at early times, upon its completion (i.e., when $\mu^{2}$ returns
to $0$), leaves no effective production. As the conformal time evolves,
$\mu^{2}(\eta)$ oscillates more and more rapidly, eventually breaking
adiabaticity with the oscillation frequency approaching $k$. From
the perspective of the physical time $t$, these background oscillations
approach a certain frequency $m_{\phi}$ while the physical momentum
$k_{{\rm phy}}=k/a$ is red-shifted. In either perspective, $k/a\sim m_{\phi}$
will be reached, leading to resonant production, as indicated by the
light red region in Fig.~\ref{fig:fk4}. In the right panel, we present
a zoom-in view of the curve, which shows that, within the resonant
region, the contributions of these oscillations are added constructively.
When $k/a\ll m_{\phi}$, these oscillations cause destructive interference
and effectively stop contributing to the production. 

\begin{figure}
\centering

\includegraphics[width=0.49\textwidth]{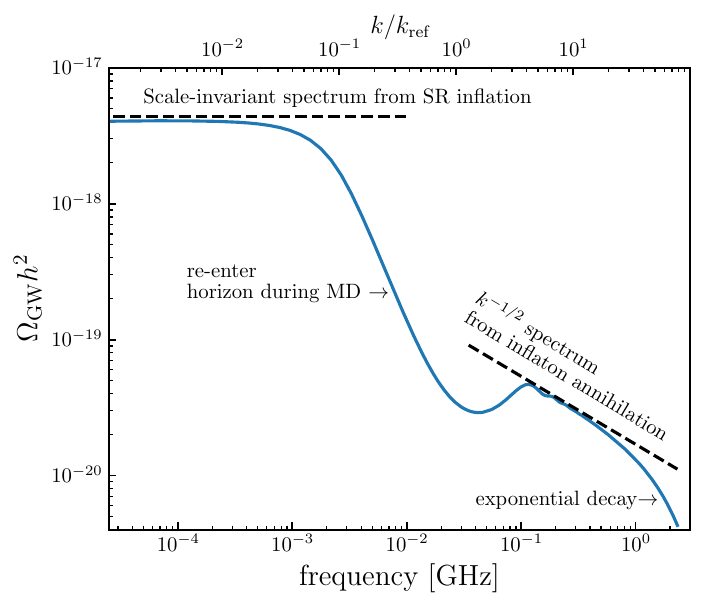}\includegraphics[width=0.498\textwidth]{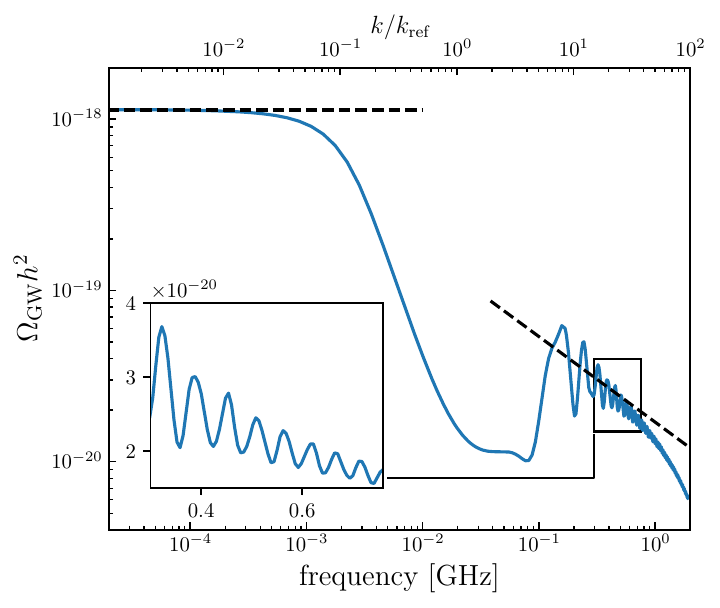}\caption{\label{fig:result} \emph{Left Panel}: GW spectrum in the T model,
computed in the unified Bogoliubov approach. The black dashed lines
represent the scale-invariant spectrum predicted by SR inflation {[}Eq.~\eqref{eq:-26}{]}
and the $k^{-1/2}$ spectrum arising from inflaton annihilation during
reheating {[}Eq.~\eqref{eq:-55}{]}, as calculated in Ref.~\cite{Choi:2024ilx}.
\emph{Right Panel}: similar to the left panel but for the Starobinsky
model. The resulting GW spectrum is more oscillatory at high frequencies
due to inflaton self-interactions---see discussions below Eqs.~\eqref{eq:-48}
and \eqref{eq:-49}. In both models, we use the same inflaton mass
$m_{\phi}=6.41\times10^{-6}M_{P}$ to determine the inflaton potential,
and the same decay rate $\Gamma_{\phi}=10^{-3}m_{\phi}$. }
\end{figure}

By varying $k$ within a few orders of magnitude around $k_{{\rm ref}}$
and solving Eq.~\eqref{eq:-13}, we obtain the full GW spectrum for
the T model (assuming $\Gamma_{\phi}=10^{-3}m_{\phi}$), as shown
in the left panel of Fig.~\ref{fig:result}.  Here $\Omega_{{\rm GW}}h^{2}$
is computed using Eq.~\eqref{eq:-17} with $f$ obtained from Eq.~\eqref{eq:-14},
and the frequency of GWs on the bottom x-axis is related to $k$ (indicated
on the top x-axis) by Eq.~\eqref{eq:-18}. 

As is expected, the resulting GW spectrum at low frequencies is approximately
scale-invariant, and it agrees well with the analytical estimate given
by Eq.~\eqref{eq:-26}, which is represented by the horizontal black
dashed line. Strictly speaking, the value of $H_{I}$ used in Eq.~\eqref{eq:-26}
should weakly depend on $k$\,---\,large-$k$ modes exit the horizon
at relatively late times with slightly smaller $H$ during SR. In
this plot, we use $H_{I}=5.66\times10^{-6}M_{P}$, corresponding to
the Hubble parameter when $k=10^{-2}k_{{\rm ref}}$ exits the horizon.
If $10^{-3}k_{{\rm ref}}$ or $10^{-1}k_{{\rm ref}}$ is used, $H_{I}/(10^{-6}M_{P})$
becomes $5.84$ or $5.22$, respectively. 

At higher frequencies (from $\sim$2 MHz to $\sim20$ MHz), the resulting
GW spectrum decreases significantly with increasing $k$. These modes
re-enter the horizon when the inflaton is oscillating at the quadratic
bottom.  Hence they re-enter the horizon in an MD epoch---see, e.g.,
\cite{Xu:2025wjq} for further discussions. Meanwhile, their frequencies
are significantly lower than the frequency of background (inflaton)
oscillations. So this part of the spectrum is not sensitive to the
details of inflaton oscillations (such as how fast the inflaton oscillates). 

At very high frequencies with $k\gtrsim k_{{\rm ref}}$, the modes
never exit the horizon and their production mainly relies on the resonance
previously illustrated in Fig.~\ref{fig:fk4}. In this regime, one
can interpret it as inflaton annihilation ($\phi\phi\to hh$ where
$h$ is a graviton), as has been recently calculated in Ref.~\cite{Choi:2024ilx}.\footnote{We note here that earlier calculations in Refs.~\cite{Ema:2015dka,Ema:2020ggo}
have already obtained very similar results, up to ${\cal O}(1)$ normalization
factors in the production rates---see Eq.~(17) in \cite{Ema:2020ggo}
versus Eq.~(11) in \cite{Choi:2024ilx}. In Ref.~\cite{Xu:2025wjq},
a similar calculation reproduces Eq.~(11) in \cite{Choi:2024ilx}.} The result of Ref.~\cite{Choi:2024ilx}, if neglecting an exponential
factor accounting for the decay of inflatons, simplifies to
\begin{equation}
\Omega_{{\rm GW}}h^{2}\approx1.7\times10^{-20}\left(\frac{\nu}{\text{GHz}}\right)^{-\frac{1}{2}}\left(\frac{g_{\star}}{106.75}\right)^{-\frac{3}{8}}\left(\frac{m_{\phi}}{6.41\times10^{-6}M_{P}}\right)^{\frac{9}{4}}\left(\frac{\Gamma_{\phi}}{10^{-3}m_{\phi}}\right)^{\frac{3}{4}}.\label{eq:-55}
\end{equation}
This spectrum scales as $k^{-1/2}\propto\nu^{-1/2}$, corresponding
to the tilted black dashed line  in Fig.~\ref{fig:result}. Further
increasing $k$, this $k^{-1/2}$ spectrum will eventually be replaced
by an exponential tail due to inflaton decays, which can be seen from
the lower right corner in the plot. 

Finally, let us note that the spectrum in the left panel of Fig.~\ref{fig:result}
exhibits small wiggles around 0.2 GHz. These wiggles are physical,
not due to numerical inaccuracies. They arise from the fact that the
inflaton oscillations are slightly anharmonic. We will see that this
effect becomes more significant in models with larger anharmonicity.

\subsection{The Starobinsky model\label{subsec:S}}

The Starobinsky model~\cite{Starobinsky:1980te} has the following
potential:
\begin{equation}
V=\frac{3}{4}M_{P}^{2}m_{\phi}^{2}\left[1-\exp\left(-\sqrt{\frac{2}{3}}\frac{\phi}{M_{P}}\right)\right]^{2},\label{eq:-56}
\end{equation}
where the parameter $m_{\phi}$ can be directly identified as the
inflaton mass when $\phi$ is small and oscillates near the minimum
of the potential. For a fair comparison with the T model, here we
fix $m_{\phi}$ at the same value as that in the T model and use the
same decay rate, $\Gamma_{\phi}=10^{-3}m_{\phi}$. 

The numerical calculation for this model is similar to that in Sec.~\ref{subsec:T}
and the resulting GW spectrum is presented in the right panel of Fig.~\ref{fig:result}.
Compared to the left panel, the most distinct feature is that the
high-frequency part starting from about $0.1$ GHz is significantly
more oscillatory. On average, this part still follows the power law
of $k^{-1/2}$ (the tilted black dashed line). The oscillations can
be understood from inflaton self-interactions stemming from the anharmonic
potential. For $\phi$ oscillating around the minimum, we expand Eq.~\eqref{eq:-56}
in $\phi$ and obtain
\begin{equation}
V\approx\frac{m_{\phi}^{2}}{2}\phi^{2}-\frac{m_{\phi}^{2}}{\sqrt{6}M_{P}}\phi^{3}+\frac{7m_{\phi}^{2}}{36M_{P}^{2}}\phi^{4}+{\cal O}\left(\phi^{5}\right),\label{eq:-48}
\end{equation}
to be compared with the expansion of $V$ in the T model:
\begin{equation}
V\approx\frac{m_{\phi}^{2}}{2}\phi^{2}-\frac{m_{\phi}^{2}}{18M_{P}^{2}}\phi^{4}+{\cal O}\left(\phi^{5}\right).\label{eq:-49}
\end{equation}
One can see that Eq.~\eqref{eq:-49} has no $\phi^{3}$ term and its
$\phi^{4}$ term is smaller than that in Eq.~\eqref{eq:-48}. Therefore,
the inflaton in the $T$ model is less self-interacting than that
in the Starobinsky model. The self-interactions, if sufficiently strong,
would invalidate the assumption that $\phi$ during the oscillation
phase can be viewed as a condensate of free bosons, which is the underlying
assumption used to derive Eq.~\eqref{eq:-55}. 

Due to the anharmonicity of the potential, the physical mass of inflatons
is actually wobbling when $\phi$ is slightly away from zero and oscillates
around it. For instance, if we use $\phi=\phi_{0}+\delta\phi$ with
$\phi_{0}=10^{-2}M_{P}$ and expand both potentials at $\phi_{0}$,
then the quadratic terms become $0.488m_{\phi}^{2}\delta\phi^{2}$
and $0.49998m_{\phi}^{2}\delta\phi^{2}$ for the Starobinsky and the
T models, respectively. Obviously, the latter is much more stable
when $\phi_{0}$ oscillates around the minimum of the potential. If
$\delta\phi$ is quantized, the particles (quanta) have a wobbling
mass $m_{\phi}(t)$, resulting in a wobbling frequency for gravitons
produced via annihilation ($\delta\phi\delta\phi\to hh$). 

If the wobbling effect is weak, such as that in the T model, each
graviton produced has a relatively steady frequency (or momentum).
And the resulting $k^{-1/2}$ spectrum is simply a consequence of
a steady production source combined with cosmological redshift. When
the wobbling effect becomes significant, like the case of the Starobinsky
model, the unsteady production combined with cosmological redshift
causes those wiggles in the right panel of Fig.~\ref{fig:result}.

\section{Conclusion \label{sec:conclusion}}

Primordial gravitational waves generated in the inflationary framework
have a rather broad spectrum. In this work, we present a unified Bogoliubov
approach that allows us to straightforwardly and effectively compute
this broad spectrum. Such a unified calculation is numerically challenging,
but with crucial insights gained from several analytical examples
(see Sec.~\ref{sec:Analytical-solutions}), we develop practically
useful techniques (see Sec.~\ref{sec:D-parametrization}) to overcome
known shortcomings in the conventional Bogoliubov approach. 

The improved method allows us to calculate the full spectrum spanning
a wide frequency range from the scale-invariant part to the high-frequency
tail without switching differential equations.  We find that anharmonicity
of inflaton oscillations during the reheating phase may leave highly
non-trivial fingerprints on the high-frequency part of the spectrum.
For instance, despite the well-known resemblance between the Starobinsky
model and a certain type of the $\alpha$-attractor T model, the former
can cause much more significant wiggles on the spectrum than the latter.
Such features could be important for probing inflationary models and
the related reheating dynamics.

\noindent\makebox[\linewidth]{\rule{0.5\linewidth}{0.4pt}}

\noindent Note added: As we were finalizing this work, we received
a preprint from the authors of Ref.~\cite{Wang:2026ule}, which contains
some overlap with this work. Ref.~\cite{Wang:2026ule} mainly addresses
the equivalence between  the Bogoliubov and Boltzmann approaches
 in describing graviton production from the inflaton condensate.
Our work is more focused on building a unified Bogoliubov approach
that can be applied to the full-spectrum calculation. Hence there
is substantial complementarity between the two papers.

\begin{acknowledgments}
We thank the authors of Ref.~\cite{Wang:2026ule} and Nicol\'as
Bernal for very useful discussions. This work is supported in part
by the National Natural Science Foundation of China under grant No.~12141501
and also by the CAS Project for Young Scientists in Basic Research
(YSBR-099). 
\end{acknowledgments}

\appendix

\section{Useful relations\label{sec:relations}}

In this appendix, we derive a few useful relations among $a$, $\eta$,
$R$, and $\mu^{2}$. We begin with  the Friedmann-Robertson-Walker
(FRW)  metric, which reads
\begin{equation}
ds^{2}=dt^{2}-a^{2}(t)\,d\mathbf{x}^{2}=a^{2}(\eta)\,(d\eta^{2}-d\mathbf{x}^{2})\thinspace,\label{eq:FRW}
\end{equation}
where  the conformal time $\eta$ is related to the cosmic time $t$
via 
\begin{equation}
d\eta=\frac{dt}{a(t)}\thinspace.\label{eq:deta-dt}
\end{equation}
With the Hubble parameter defined as $H=\dot{a}/a$,  we express
$d\eta$ in terms of $da$ as 
\begin{equation}
d\eta=\frac{da}{a}\frac{dt}{da}=\frac{da}{a^{2}H}\thinspace.\label{eq:deta-da}
\end{equation}

Integrating both sides of Eq.~(\ref{eq:deta-da}) gives the relation
between $\eta$ and $a$, if the dependence of $H$ on $a$ is known.
Since $H\propto\rho^{1/2}$, this depends on how $\rho$ scales with
$a$. It is well known that during slow-roll (SR), matter domination
(MD), and radiation domination (RD), $\rho$ scales as $a^{0}$, $a^{-3}$,
and $a^{-4}$, respectively. More generally, if the universe is dominated
by a certain form of energy with the equation of state parameter $w\equiv p/\rho$
being constant, then we have $\rho\propto a^{-3\,(1+w)}$ and therefore
\begin{equation}
H=H_{1}\left(\frac{a}{a_{1}}\right)^{-\frac{3\,(1+w)}{2}},\label{eq:H-scaling}
\end{equation}
where $a_{1}$ and $H_{1}$ are generic reference values within that
epoch. 

Substituting Eq.~(\ref{eq:H-scaling}) into Eq.~(\ref{eq:deta-da})
and integrating both sides of the equation, we obtain
\begin{equation}
\eta-C_{\eta}=\frac{2}{(1+3w)\,a_{1}H_{1}}\left(\frac{a}{a_{1}}\right)^{\frac{1+3w}{2}},\label{eq:eta-a}
\end{equation}
or, equivalently, 
\begin{equation}
a=a_{1}\left[\frac{(1+3w)\,a_{1}H_{1}}{2}\,(\eta-C_{\eta})\right]^{\frac{2}{1+3w}}.\label{eq:-65}
\end{equation}
Here $C_{\eta}$ is an unfixed constant. For MD ($w=0$) and RD ($w=1/3$),
the right-hand side of Eq.~(\ref{eq:eta-a}) vanishes at $a\to0$;
and for SR ($w=-1$), it vanishes at $a\to\infty$. For convenience,
$\eta=0$ in the literature is often set at the vanishing point of
the right-hand side, which would fix $C_{\eta}$ at zero. However,
one needs to be careful with this setting when multiple epochs (e.g.,
SR+MD+RD+MD) are connected. In such cases, not all epochs are allowed
to consistently fix their respective $C_{\eta}$ at zero. Hence in
this appendix, we keep $C_{\eta}$ unfixed.

Next, we proceed with the calculation of the Ricci scalar $R$, starting
from the Christoffel symbols defined as $\tensor{\Gamma}{_{\mu\nu}^{\rho}}=\frac{1}{2}g^{\rho\lambda}\,(\partial_{\mu}g_{\nu\lambda}+\partial_{\nu}g_{\mu\lambda}-\partial_{\lambda}g_{\mu\nu})$.
With Eq.~\eqref{eq:FRW},  one can compute $\tensor{\Gamma}{_{\mu\nu}^{\rho}}$
straightforwardly and find the non-vanishing components  to be 
\begin{align}
\tensor{\Gamma}{_{ij}^{0}} & =a\,\dot{a}\,\delta_{ij}=Ha^{2}\delta_{ij}\thinspace,\\
\tensor{\Gamma}{_{0j}^{i}}=\tensor{\Gamma}{_{j0}^{i}} & =\frac{\dot{a}}{a}\,\tensor{\delta}{_{j}^{i}}=H\tensor{\delta}{_{j}^{i}}\thinspace,
\end{align}
where Latin indices $i$, $j$ denote spatial coordinates. The Riemann
curvature tensor is defined as $\tensor{R}{_{\sigma\mu\nu}^{\rho}}=\partial_{\mu}\tensor{\Gamma}{_{\nu\sigma}^{\rho}}-\partial_{\nu}\tensor{\Gamma}{_{\mu\sigma}^{\rho}}+\tensor{\Gamma}{_{\mu\lambda}^{\rho}}\tensor{\Gamma}{_{\nu\sigma}^{\lambda}}-\tensor{\Gamma}{_{\nu\lambda}^{\rho}}\tensor{\Gamma}{_{\mu\sigma}^{\lambda}}$,
whose non-zero components in the FRW spacetime are listed as follows:
\begin{align}
\tensor{R}{_{i0j}^{0}} & =a\,\ddot{a}\,\delta_{ij}=\left(\dot{H}+H^{2}\right)\,a^{2}\,\delta_{ij}\thinspace,\\
\tensor{R}{_{jkl}^{i}} & =\dot{a}^{2}\left(\tensor{\delta}{_{k}^{i}}\delta_{jl}-\tensor{\delta}{_{l}^{i}}\delta_{jk}\right)=H^{2}a^{2}\left(\tensor{\delta}{_{k}^{i}}\delta_{jl}-\tensor{\delta}{_{l}^{i}}\delta_{jk}\right).
\end{align}
Contracting the indices to obtain the Ricci tensor $R_{\mu\nu}=\tensor{R}{_{\mu\lambda\nu}^{\lambda}}$,
we have 
\begin{align}
R_{00} & =-\frac{3\,\ddot{a}}{a}=-3\left(\dot{H}+H^{2}\right),\\
R_{ij} & =(\dot{H}+3H^{2})\,a^{2}\,\delta_{ij}=\left(a\,\ddot{a}+2\,\dot{a}^{2}\right)\delta_{ij}\thinspace.
\end{align}
From the trace of the Ricci tensor, we obtain the Ricci scalar:
\begin{align}
R & =\tensor{R}{_{\mu}^{\mu}}=g^{\mu\nu}R_{\mu\nu}\nonumber \\
 & =-\frac{6}{a^{2}}\left(\ddot{a}\,a+\dot{a}^{2}\right)=-6\left(\dot{H}+2H^{2}\right),
\end{align}
or, in terms of conformal time derivatives,
\begin{equation}
R=-\frac{6\,a''}{a^{3}}\thinspace.
\end{equation}

The background energy density $\rho$ evolves according to the continuity
equation as 
\begin{equation}
\dot{\rho}+3H(\rho+p)=0\thinspace.
\end{equation}
Combining this with $\rho=3M_{P}^{2}H^{2}$ and Eq.~\eqref{eq:H-scaling},
we obtain  $\dot{H}=-3\,(1+w)H^{2}/2$, which allows us to express
the Ricci scalar as 
\begin{equation}
R=-3\,(1-3w)H^{2}\thinspace.\label{eq:-63}
\end{equation}

Alternatively, using the trace of the Einstein field equations, 
one can directly relate $R$ to the matter sector via 
\begin{equation}
R=\frac{1}{M_{P}^{2}}\sum_{i}(3\,p_{i}-\rho_{i})\thinspace,\label{eq:-64}
\end{equation}
which is equivalent to Eq.~\eqref{eq:-63}. 

For a homogeneous, dynamical scalar field $\phi$, its energy density
and pressure are given by 
\begin{equation}
\rho_{\phi}=\frac{1}{2}\dot{\phi}^{2}+V\thinspace,\quad p_{\phi}=\frac{1}{2}\dot{\phi}^{2}-V\thinspace.
\end{equation}
Substituting these into Eq.~\eqref{eq:-64}, we find that its contribution
to the Ricci scalar is 
\begin{equation}
R_{\phi}=\frac{\dot{\phi}^{2}-4V}{M_{P}^{2}}\thinspace.
\end{equation}

Finally, let us compute $\mu^{2}\equiv a''/a=-a^{2}R/6$.  This can
be done using Eq.~\eqref{eq:-63} or ~\eqref{eq:-64}, or by directly
computing derivatives of Eq.~\eqref{eq:-65} with respect to $\eta$.
The result is
\begin{equation}
\mu^{2}=\frac{1-3w}{(1+3w)^{2}}\,\frac{2}{(\eta-C_{\eta})^{2}}\thinspace.
\end{equation}

For a few commonly considered values of $w$, we present the relations
of $\eta$ vs $a$, $R$ vs $\eta$, and $\mu^{2}$ vs $\eta$ explicitly
as follows: 
\begin{itemize}
\item \textbf{SR} ($w=-1$):
\begin{equation}
\eta-C_{\eta}=-\frac{1}{aH}\propto-\frac{1}{a}\thinspace,\quad R=-12H^{2}\thinspace,\quad\mu^{2}=\frac{2}{(\eta-C_{\eta})^{2}}\thinspace.
\end{equation}
\item \textbf{MD} ($w=0$): 
\begin{equation}
\eta-C_{\eta}=\frac{2}{aH}\propto\sqrt{a}\thinspace,\quad R=-3H^{2}\propto\frac{1}{(\eta-C_{\eta})^{6}}\thinspace,\quad\mu^{2}=\frac{2}{(\eta-C_{\eta})^{2}}\thinspace.
\end{equation}
\item \textbf{RD} ($w=1/3$): 
\begin{equation}
\eta-C_{\eta}=\frac{1}{aH}\propto a\thinspace,\quad R=0\thinspace,\quad\mu^{2}=0\thinspace.
\end{equation}
\end{itemize}

\section{Suppression of UV divergences\label{sec:Suppression-of-UV}}

In Sec.~\ref{sec:Analytical-solutions}, we have seen that any sharp
cut-off would lead to UV divergences in the resulting GW energy density
(i.e., $k^{4}f(k)$ does not vanish for $k\to\infty$). If the sharp
cut-off is replaced by a continuous transition taking finite time,
the UV divergence should disappear. In particular, according to Eq.~\eqref{eq:-45},
we expect that $k^{4}f(k)$ in the large-$k$ limit should be suppressed
when the adiabaticity parameter $|A_{k}|\ll\mu^{2}/k^{2}$. In this
appendix, we present a concrete example to numerically demonstrate
this.

\begin{figure}
\centering

\includegraphics[width=0.85\textwidth]{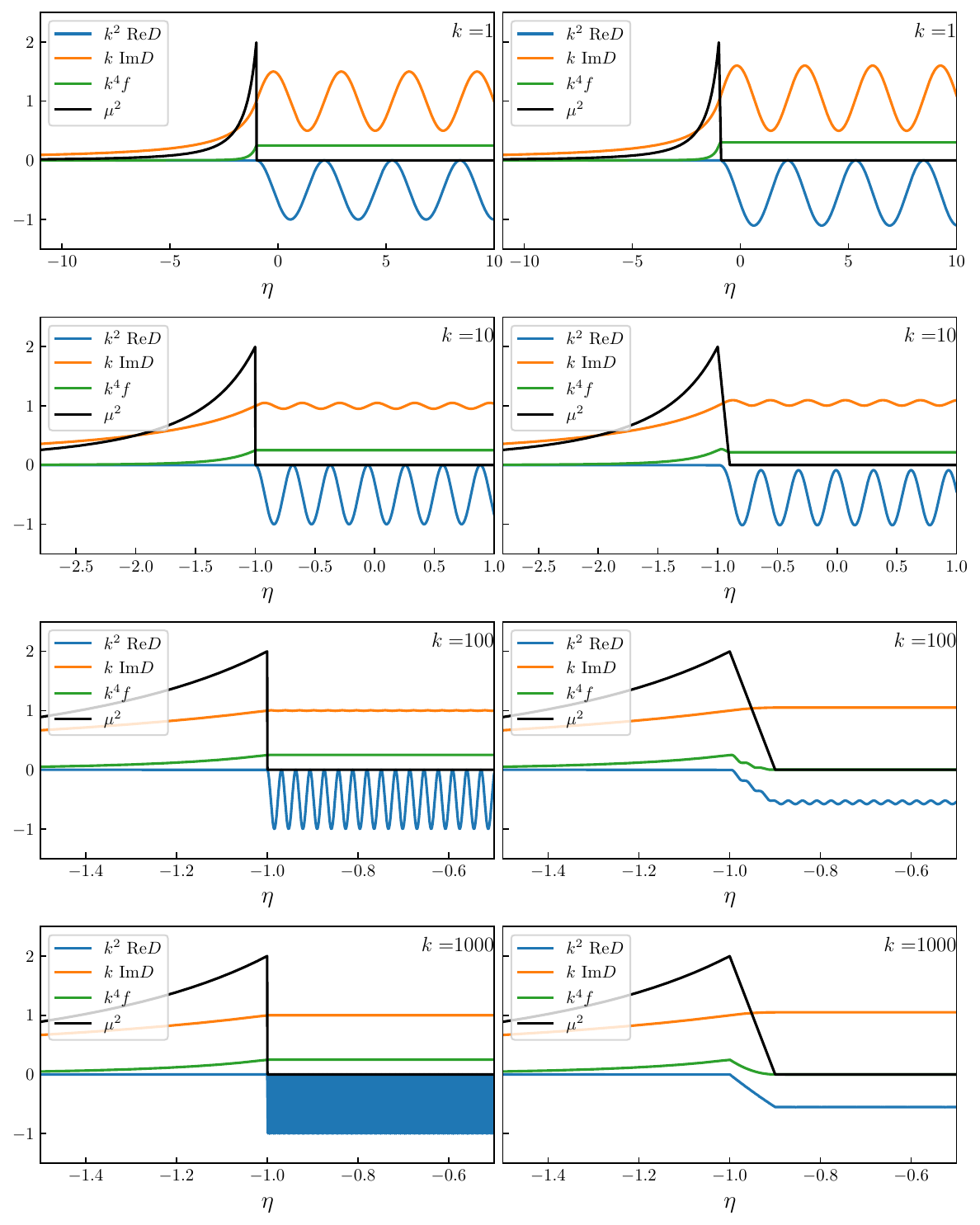}

\caption{\label{fig:UV} \emph{Left panels}: Numerical solutions of Eq.~\eqref{eq:-13}
with $\mu^{2}$ given by Eq.~\eqref{eq:-24-1} and $\delta\eta=0$.
\emph{Right panels}: Similar plots but with $\delta\eta=0.1$  (see
also difference between the black lines on the left and right panels). }
\end{figure}

We take the $\mu^{2}$ function in Eq.~\eqref{eq:-24} and modify
it as follows:
\begin{equation}
\mu^{2}\left(\eta\right)=\begin{cases}
\frac{2}{\eta^{2}} & \eta\leq\eta_{1}\\
\frac{2}{\eta_{1}^{2}}\left(1-\frac{\eta-\eta_{1}}{\delta\eta}\right) & \eta_{1}<\eta\leq\eta_{1}+\delta\eta\\
0 & \eta>\eta_{1}+\delta\eta
\end{cases}\thinspace.\label{eq:-24-1}
\end{equation}
This avoids a sharp cut-off at $\eta=\eta_{1}$ and lets $\mu^{2}$
linearly decrease to zero after $\eta_{1}$. With $\eta_{1}=-1$ and
$\delta\eta=0.1$ or $0$, we solve Eq.~\eqref{eq:-13} numerically
and present the results in Fig.~\ref{fig:UV}. 

From Fig.~\ref{fig:UV}, one can see that for $k=1$ the solutions
and the resulting values of $k^{4}f$ are almost unaffected by the
small difference caused by $\delta\eta$. As $k$ increases to larger
values, the final values of $k^{4}f$ in the left panels remain unchanged,
but in the right panels they are suppressed for large $k$. By imposing
the condition $|A_{k}|\ll\mu^{2}/k^{2}$ on the linearly decreasing
part, we  obtain $2/(\eta_{1}^{2}k^{3}\delta\eta)\ll2/(k^{2}\eta_{1}^{2})$,
corresponding to $k\gg1/\delta\eta=10$. Indeed, we see in Fig.~\ref{fig:UV}
that for $k=100$ and $1000$, the final values of $k^{4}f$ are suppressed
to approximately zero.

\section{The adiabaticity parameter for UV smoothing\label{sec:adiabatic}}

When the UV smoothing function in Eq.~\eqref{eq:-47} is used during
slow-roll, the corresponding adiabaticity parameter $A_{k}$ can be
estimated analytically. Substituting Eq.~\eqref{eq:-47} into Eq.~\eqref{eq:-46}
and assuming $k^{2}\gg\mu^{2}$, we obtain
\begin{equation}
A_{k}\approx\exp\left[-\epsilon\left(\frac{\eta_{b}-\eta_{a}}{\eta-\eta_{a}}\right)^{2}\right]\frac{2}{k^{2}\eta^{2}}\left(\frac{\epsilon}{k(\eta-\eta_{a})}\left(\frac{\eta_{b}-\eta_{a}}{\eta-\eta_{a}}\right)^{2}-\frac{1}{k\eta}\right).\label{eq:-67}
\end{equation}

Comparing $A_{k}$ with $\mu^{2}/k^{2}\approx2/(k^{2}\eta^{2})$,
we see that the strong adiabaticity condition $A_{k}\ll\mu^{2}/k^{2}$
corresponds to 
\begin{equation}
\exp\left[-\epsilon\left(\frac{\eta_{b}-\eta_{a}}{\eta-\eta_{a}}\right)^{2}\right]\left(\frac{\epsilon}{k(\eta-\eta_{a})}\left(\frac{\eta_{b}-\eta_{a}}{\eta-\eta_{a}}\right)^{2}-\frac{1}{k\eta}\right)\ll1\thinspace.\label{eq:-68}
\end{equation}
The last factor $\frac{1}{k\eta}$ can be neglected because $k|\eta|\gg1$
and the exponential factor is always less than one. Dropping this
term, one can compute the maximum of the left-hand side analytically.
The maximum is reached when 
\begin{equation}
\epsilon\left(\frac{\eta_{b}-\eta_{a}}{\eta-\eta_{a}}\right)^{2}=\frac{3}{2}\thinspace.\label{eq:-69}
\end{equation}
Substituting it back into Eq.~\eqref{eq:-68}, we obtain
\begin{equation}
\sqrt{\frac{27}{8\epsilon}}\frac{e^{-3/2}}{k(\eta_{b}-\eta_{a})}\ll1\thinspace.\label{eq:-70}
\end{equation}
It implies that for the UV smoothing part to satisfy the strong adiabaticity
condition, $\epsilon$ and $|\eta_{b}-\eta_{a}|$ should not be too
small. 

In addition to Eq.~\eqref{eq:-47}, we provide an alternative form
for UV smoothing, with the following smoothing function
\begin{equation}
W(\eta)=\left\{ \begin{array}{ll}
0\thinspace, & x\le0\thinspace,\\[0.3em]
\left[1+\exp\left(\dfrac{1}{x}-\dfrac{1}{1-x}\right)\right]{}^{-1}\thinspace,\ \  & 0<x<1\thinspace,\\[0.8em]
1\thinspace, & x\ge1\thinspace,
\end{array}\right.\label{eq:-72}
\end{equation}
where 
\begin{equation}
x\equiv\frac{\eta-\eta_{a}}{\eta_{b}-\eta_{a}}\ \ \text{with}\ \ \Delta\eta=\eta_{b}-\eta_{a}>0\thinspace.\label{eq:-71}
\end{equation}
Eq.~\eqref{eq:-72} satisfies $W(\eta)\in C^{\infty}(\mathbb{R})$,
$W(\eta_{a})=0$, and $W(\eta_{b})=1$ exactly. The smoothing is then
implemented as $\mu^{2}(\eta)\to W(\eta)\,\mu^{2}(\eta)$.

Under the assumption that $\mu^{2}$ before being processed by $W$
satisfies the strong adiabaticity condition, the adiabaticity parameter
is approximately determined by $A_{k}\approx\mu^{2}W'(\eta)/(2k^{3})$.
Hence $A_{k}\ll\mu^{2}/k^{2}$ corresponds to $W'(\eta)\ll2k$. Analytically,
one can find that $W'(\eta)$ reaches its maximum, $2/\Delta\eta$,
when $x=1/2$. In order to maintain $A_{k}\ll\mu^{2}/k^{2}$, we require
$2/\Delta\eta\ll2k$, i.e., 
\begin{equation}
k\Delta\eta\gg1\thinspace,\label{eq:-73}
\end{equation}
which is the the condition for the UV smoothing function \eqref{eq:-72}
to satisfy the strong adiabaticity condition. 

\bibliographystyle{JHEP}
\bibliography{ref}

\end{document}